\documentclass[nohyper]{JHEP3}

\usepackage{cite}
\usepackage{epsfig}
\usepackage{amsmath}

\newcommand{\secn}[1]{Section~\ref{#1}}

\newcommand{\ee}{\eeq}

\def\beq{\begin{equation}}
\def\eeq{\end{equation}}
\def\beqa{\begin{eqnarray}}
\def\eeqa{\end{eqnarray}}
\def\eq#1{Eq.~(\ref{#1})}

\def\PT{\hbox{\tiny PT}}

\def\NP{\hbox{\tiny NP}}

\newcommand{\alg}{\mathrm{alg}}

\newcommand{\cone}{\mathrm{cone}}
\newcommand{\rf}{\mathrm{ref}}
\newcommand{\alt}{\mathrm{alt}}
\newcommand{\NLO}{\mathrm{NLO}}

\newcommand{\ie}{\emph{i.e.}\ }

\def\order#1{{\cal{O}}\left(#1\right)}

\newcommand{\GeV}{\,\mathrm{GeV}}
\newcommand{\TeV}{\,\mathrm{TeV}}
\newcommand{\ps}{\mathrm{ps}}

\newcommand{\h}{\mathrm{h}}
\newcommand{\UE}{\mathrm{UE}}

\def\MSbar{\overline{\mbox{\scriptsize MS}}}

\def\CF{C_F}
\def\TR{T_R}
\def\CA{C_A}

\def\nf{n_{\!f}}
\def\as{\alpha_s}

\def\ee{e^+e^-}

\title{Non-perturbative QCD effects in jets at hadron colliders}

\author{Mrinal Dasgupta\\
School of Physics and Astronomy, University of Manchester\\
Oxford Road, Manchester M13 9PL, U.K.\\
E-mail: {\tt dasgupta@hep.man.ac.uk}}
\author{Lorenzo Magnea \\
Dipartimento di Fisica Teorica, Universit{\`a} di Torino, and\\
INFN, Sezione di Torino, Via P. Giuria, I--10125 Torino, Italy\\
E-mail: {\tt magnea@to.infn.it}}
\author{Gavin P. Salam\\
LPTHE, CNRS UMR 7589; Universit\'e Pierre et Marie Curie (Paris VI);
Universit\'e Denis Diderot (Paris VII), 75252 Paris Cedex 05, France\\
E-mail: {\tt salam@lpthe.jussieu.fr}}
  
\abstract{We discuss non-perturbative QCD contributions to jet observables,
computing their dependence on the jet radius $R$, and on the colour and 
transverse momentum of the parton initiating the jet. We show, using
analytic QCD models of power corrections as well as Monte Carlo simulations,
that hadronisation corrections grow at small values of $R$, behaving as $1/R$,
while underlying event contributions grow with the jet area as $R^2$.
We highlight the connection between hadronisation corrections to jets and those
for event shapes in $e^+e^-$ and DIS; we note the limited dependence of our
results on the choice of jet algorithm; finally, we propose several measurements
in the context of which  to test or implement our predictions. The results presented here reinforce
the motivation for the use of a range of $R$ values, as well as a plurality of 
infrared-safe jet algorithms, in precision jet studies at hadron colliders.}


\preprint{
  DFTT--27/2007\\
  MAN/HEP/2007/41\\
}

\begin{document}

\section{Introduction}
\label{sec:intro}

Jets play a key role in many experimental studies at the Tevatron, and
will continue to do so at the LHC. They are useful, for example, in top reconstruction and mass measurements, in searches for the Higgs boson 
and new physics signals, and of course they are instrumental for QCD studies,
such as inclusive-jet measurements, which in turn are an important input for
the determination of parton distribution functions.

Jets, however, are fundamentally ambiguous objects, reflecting the fact 
that the divergences of QCD perturbation theory make it impossible to
define clusters of final state hadrons uniquely assigned to individual hard 
partons emitted at leading order. Most procedures to cluster measured
hadrons into jets, \ie jet algorithms, deal with this ambiguity by introducing 
a resolution parameter to define when exactly a jet with substructure should
actually be resolved into two jets. For currently used hadron-collider jet 
algorithms the main parameter is generally a threshold on the allowed 
opening angle of the jet, and is called $R$ or jet radius (defined on the 
azimuth-rapidity cylinder).

Given, on the experimental side, the increased precision required for present
hadron collider studies, and the complex hadronic environment expected in 
LHC final states, and given many years of theoretical advances in our understanding of QCD,  it is both useful and possible to work towards
a better quantitative understanding of how the choice of the jet algorithm 
and the specification of its parameters influence the clustering of QCD radiation,
and our general understanding of the final state in hadron collisions. 

One can consider three classes of QCD effects: perturbative radiation,
hadronisation and the underlying event. It should be realised that there 
is no unique, fully consistent, gauge-invariant way to distinguish these 
three classes: rather, one must try to define perturbation theory, and the
underlying factorisation scheme, precisely enough, in order to identify
non-perturbative hadronisation corrections without double counting; 
underlying event corrections will, at some level, inevitably mix with 
hadronisation: one must argue that this mixing does not occur, or is 
parametrically controllable, at least at leading and next-to-leading power 
in the hard scale. In this respect, as we will see, the radius dependence is 
a very useful tool.  With this premise, one would like to understand how 
each of these classes of contributions affects the momentum of a jet, 
as a function of the jet radius $R$, as a function of the centre-of-mass 
energy, and as a function of the properties of the parton that initiated 
the jet, such as its colour charge and its transverse momentum. 

These issues can be studied through parton-shower Monte Carlo 
generators, taking advantage of their detailed and well tested 
non-perturbative modelling. It can, however,  be difficult to extract 
simple analytical understanding from such models. An alternative 
approach is to carry out analytical calculations directly. Perturbatively, 
one may work with soft and collinear approximations for gluon emission, 
and perform a resummation. The tools for this task, in the intricate
case of jet production in hadron-hadron collisions, have been available for 
a long time~\cite{Kidonakis:1997gm,Kidonakis:1998bk,Kidonakis:1998nf}, 
and phenomenological studies of the effects of soft and collinear logarithms 
near threshold have been performed, both for inclusive jet cross 
sections~\cite{Kidonakis:2000gi,deFlorian:2007fv} and for jet 
shapes~\cite{Seymour:1997kj}. It is well understood that these resummations
carry nontrivial information on the parametric size of non-perturbative  
corrections, as discussed in \cite{Korchemsky:1994is}. Furthermore, one 
can carry out non-perturbative studies using renormalon and related
approaches~\cite{Dokshitzer:1995zt,Ball:1995ni,Dokshitzer:1995qm,Gardi:2001di};
these suggest that the dynamics of Monte Carlo models can in some cases 
be reduced, for sufficiently inclusive observables, to a single
non-perturbative parameter, multiplied by analytically calculable
coefficients. These models have had considerable success in the context of 
$e^+e^-$ and DIS (for reviews, see \cite{Beneke:1998ui,Dasgupta:2003iq});
in contrast, for QCD final-state observables at hadron colliders, only a handful 
of results exist, notably for the jet energy-flow profile \cite{Seymour:1997kj}, 
the out-of-plane momentum for $Z+$jet production \cite{Banfi:2001aq} and 
away-from-jet energy flow~\cite{Dasgupta:2007hr}.

It is perhaps worth emphasising that the non-perturbative corrections we 
are discussing, especially at moderate $p_t$, have an impact that is 
comparable to that of higher-order perturbative effects, and may affect in 
a crucial way precision studies for many observables, even in the nominally
asymptotic energy regime probed by the LHC. This is a familiar issue from 
QCD studies at LEP and HERA, especially those concerning event shapes. 
In the context of hadron collider phenomenology, non-perturbative radiation 
from hadronisation and underlying event is, for example, the main source of
uncertainty in the determination of the jet energy scale. This plays a key role 
in precision studies, as was shown at Tevatron~\cite{Bhatti:2005ai}: each
percentage point of uncertainty in the jet energy scale translates, for example, 
into a 1 GeV uncertainty in the top mass, and into a $10 \%$ uncertainty in 
the single-inclusive jet $p_t$ distribution at $p_t \sim 500$ GeV. In general, 
the steep slope of jet distributions amplifies the effects of comparatively 
small energy shifts due to non-perturbative radiation, and propagates their 
impact all the way to the highest energies~\cite{Mangano:1999sz}: this 
effect will actually be magnified at LHC, where the underlying event is 
expected to provide an energy density much higher than that measured 
at Tevatron.

In this paper, we begin to address quantitatively a number of issues related 
to non-perturbative corrections to jet observables at hadron colliders. We focus 
mostly on the radius dependence of non-perturbative effects, but examine
also the dependence on the quantum numbers of the partons initiating the
jets, and the effects of the choice of jet algorithm, thus complementing and
extending the more qualitative statements found for example in
\cite{Campbell:2006wx,Ellis:2007ib}. For the sake of completeness, and in 
order to be able to compare perturbative and non-perturbative effects as 
precisely as possible, we begin in \secn{sec:pertR} by reviewing and organising
perturbative results at NLO. Similar results have been derived and used before
in order, for example, to allow for a precise matching between resummed and 
NLO jet cross sections~\cite{deFlorian:2007fv}. In \secn{sec:analytical}, we 
move to the main subject of our analysis, and provide an analytical estimate 
for the change $\delta p_t$ in the transverse momentum of a jet due to
hadronisation, at leading power in $p_t$. We employ the techniques of the
dispersive approach to power corrections, adapted to the environment of hadron
collisions. Our main result concerns the radius dependence of $\delta p_t$, which
behaves like $1/R$ at small $R$, in sharp contrast to the behaviour of
underlying event contributions, which are expected to grow with the jet area 
and become negligible for very narrow jets.  The $1/R$ dependence of
$\delta p_t$ could have been deduced from the analysis 
of~\cite{Korchemsky:1994is}, predicting a $1/(Q R)$ correction to the
jet cross section. This fact seems however not to have been widely appreciated.
Relative to~\cite{Korchemsky:1994is}, here we work with fully-defined jet
algorithms, we show that corrections of $\order{1}$ vanish, while those
of $\order{R}$ have a small coefficient, we provide the relation between the
coefficient of $1/R$ and the non-perturbative parameter $\alpha_0$ used in 
event shape studies (cf.\ \cite{Dasgupta:2003iq} and references therein),
separately for quark- and gluon-induced jets, and we outline the class of jet
algorithms for which this relation is expected to be exact.

In \secn{sec:montecarlo} we compare our analytical results with Monte Carlo
simulations, while varying $R$, the main parameter entering the jet definition. 
We employ {\tt Pythia}~\cite{Sjostrand:2006za}, as well as 
{\tt Herwig}~\cite{Corcella:2002jc} (with {\tt Jimmy}~\cite{Butterworth:1996zw}
providing the underlying event) to generate hadron level jet events, and we
reconstruct the jets using different (IR safe) algorithms, for different parton
channels and different centre of mass energies. It is possible to generate 
separately the hadronisation and underlying event components of $\delta p_t$
with Monte Carlo simulations; the main result for the hadronisation contribution 
is a striking confirmation of our analytic expression: the four different jet 
algorithms that we employ all agree with each other and with the analytic 
result, both in shape and normalisation, within margins that are quite 
reasonable given the inherent approximations of both models. The underlying 
event component is certainly much less well understood; it is clear, in any case,
that the momentum shift due to the underlying event grows with $R$, and for
moderate values of $R$ it is several times larger at LHC than at the Tevatron, 
as might be expected.

We conclude, in \secn{sec:exp-considerations}, by outlining some experimental consequences of our results and discussing some possible measurements that
could validate them. Specifically, we discuss the optimal choice of the jet
radius $R$ in different experimental circumstances, a practical way to measure 
the $p_t$ shift due to hadronisation and underlying event, and the effect of 
such a shift on the single-inclusive jet $p_t$ distribution.

We regard these results as a first step towards a detailed analysis of jet physics
beyond fixed order perturbation theory, which could be improved upon in several
ways: our analytic estimate could for example be made more precise by matching
it to a resummed prediction for a specific jet distribution; in general, we expect 
that beyond the leading $1/R$ behaviour non-perturbative radiation will not simply
shift the cross section but also change its shape, as seen for example 
in~\cite{Dasgupta:2007hr}. One also expects that the shift will not be 
universal across different observables and algorithms, since typical jet 
algorithms introduce nonlinearities that are likely to spoil the simple pattern 
of exponentiation that underlies the power of one-gluon results~\cite{Lee:2006nr}: 
the effects of this breaking of universality could be studied both with analytical 
and Monte Carlo tools.

We believe, in any case, that our results strongly suggest that experimental
collaborations should try to maximise the flexibility  of their choices concerning 
jet analyses. A wealth of information, and perhaps even opportunities for 
discovery, can be missed, if future jet studies are confined to just one 
or two algorithms, and a handful of parameter sets, chosen ahead of time. 
LHC is a discovery machine, and in order to fully exploit its potential we must
be prepared with a range of  flexible tools, capable of meeting unexpected, as 
well as expected challenges.

\section{Perturbative $\boldsymbol{R}$ dependence}
\label{sec:pertR}

We begin by reviewing the perturbative dependence on the jet radius $R$, for 
small values of $R$, where a priori one can expect a logarithmic enhancement
originating from the collinear singularity that is approached when the jet
becomes very narrow. Specifically, we observe that perturbative $\ln R$ terms 
(discussed on many occasions previously,
\cite{Furman:1981kf,Aversa:1988vb,Guillet:1990ez,Jager:2004jh,deFlorian:2007fv})
originate from 
the same phenomenon as the $1/R$ growth that we will find for
hadronisation contributions: as the jet becomes narrow, partons
radiated 
outside of it (in other words partons which are not recombined with the jet 
by the chosen jet algorithm) are allowed to become more and more collinear
to the emitter, approaching the collinear-singular configuration. Since these 
contributions are uniquely associated with the outgoing jet, and are
independent of the other hard emitters, they can be simply computed in the
collinear approximation. 

In order to perform explicit calculations, we need to pick specific observables.
Let us first consider the loss of transverse momentum for a leading jet, which is 
of relevance for example when using the hardest pair of jets in an event to reconstruct the mass of a decayed massive particle. As an example, consider 
the quasi-collinear branching of a quark with transverse momentum $p_t$, 
which splits into a quark carrying a fraction $z$ of the initial momentum, plus 
a gluon carrying the remaining fraction $1 - z$. Given that the probability of 
such a quasi-collinear branching is just the corresponding splitting function, 
one can easily write down an expression for the leading perturbative contribution,
at small $R$, to the average change in transverse momentum of a high 
$p_t$ jet. One finds
\beq
  \label{eq:deltapt_pert_smallR}
  \langle \delta p_t \rangle_{\mathrm{pert}} = p_t \int \frac{d \theta^2} 
  {\theta^2} \int dz \big( \max[z, 1 - z] - 1 \big) \,
  \frac{\as \big(\theta \, z (1 - z) \, p_t \big)}{2 \pi} \, P_{qq} (z) \,
  \Theta \big( \theta - f_\alg(z) R \big) \, .
\eeq
Here we constructed the change in transverse momentum as the
difference between the $p_t$ of the leading jet (which can be a quark
or a gluon jet, depending on whether $z$ or $1 - z$ is larger) and
that of the initial quark, accounting for the splitting probability
$P_{qq} (z)$, averaged over the essentially collinear branching. The
$\Theta$ function constraint denotes the condition for
non-recombination of the softer parton into the leading jet, and
$f_\alg (z)$ is a function that depends on the jet algorithm. In the
$k_t$ \cite{inclukt,ESkt} and Cambridge/Aachen \cite{Caachen} algorithms
(for detailed definitions, see section~\ref{sec:montecarlo}), for
example, one merely requires that the small angle $\theta$ between the
quark and gluon be greater than $R$, so that $f_{k_t} (z) = f_{\rm
Cam} (z) = 1$. For stable-cone based algorithms, such as SISCone
\cite{Salam:2007xv}\footnote{At this order the Midpoint algorithm
  \cite{Blazey} behaves in the same way, however it suffers from infrared
  unsafety at higher orders.}, one is required, in order to enforce stability, 
to first construct the energy-weighted centroid of the quark-gluon system, 
$\vec{n}_j = z \hat{n}_q + (1 - z) \hat{n}_g$, where $\hat{n}_{q,g}$ are 
unit vectors along the the quark and gluon directions.  One must then ensure 
that the angle between $\vec{n}_j$ and the softer parton be greater than
$R$, which implies that the parton will not be recombined. Projecting, for 
example, the unit vector along the jet $\hat{n}_j = \vec{n}_j/|\vec{n}_j|$ 
onto the gluon direction, in the small-angle limit one finds
\beq
 \label{thethe}
 \theta_{j g} = z \, \theta _{q g}~,
\eeq
relating the opening angle between the gluon and jet directions, $\theta_{j g}$,
to the quark-gluon opening angle $\theta_{q g} = \theta$. Non-recombination,
for the case where the gluon is the softer parton, so that $z > 1/2$, requires 
$\theta_{j g} >R$, which in turn gives 
\beq
 \label{theqglim}
 \theta_{q g} > \frac{R}{z} = R \left (1 + \frac{1 - z}{z} \right) \, ;
\eeq
if on the other hand the quark is softer ($z < 1/2$), the corresponding result is
obtained by replacing $z$ with $1 - z$ in \eq{theqglim}. This leads to 
\beq
 \label{eq:falg}
 f_\cone (z) = 1 + \min \left( \frac{z}{1 - z}, \frac{1 - z}{z} \right) \, .
\eeq
As a consequence, for cone algorithms, the $\theta$ integral in 
\eq{eq:deltapt_pert_smallR} is cut off at $R/z$ ($R/(1 - z)$) for $z > 1/2$
($z < 1/2$). The $z$-dependence of the integration region does not affect 
the coefficient of $\ln R$, but gives an $R$-independent shift, computed below.
Note that this result is independent of the overlap threshold in the split--merge
stage of SISCone and similar algorithms.

Let us now return to \eq{eq:deltapt_pert_smallR}. The logarithmic behaviour 
comes from the $\theta$ integral, which is cut off by the $\Theta$ function 
at small angle, while the upper limit is given by the large-angle hard 
partonic structure of the event, and does not affect the $R$ dependence in 
the present approximation. For a quark-initiated jet, define then
\beq
  \label{eq:Lq}
  L_q \equiv \int d z \, \min \big(z, 1 - z \big) \,
  P_{qq} (z) = \CF \left(2 \ln 2 - \frac{3}{8} \right) \, ,
\eeq
while for a gluon-initiated jet the corresponding quantity is
\beq
  \label{eq:Lg}
  L_g \equiv  \int d z \min \big(z, 1 - z \big) \left(
  \frac{1}{2} P_{g g} (z) + n_f  P_{q g} \right) = 
  \CA \left( 2 \ln 2 - \frac{43}{96} \right) + \nf \, \TR \, \frac{7}{48} \, ,
\eeq
where we have included a $1/2$ factor for Bose symmetry in the $gg$
channel, and $P_{i j} (z)$ are the real emission parts of the leading order 
DGLAP splitting kernels. It is easy to verify that, for a jet originated by a 
parton of type $i$, \eq{eq:deltapt_pert_smallR} yields
\beq
  \label{eq:deltapt_pert_smallR_res}
  \frac{\langle \delta p_t \rangle_{\rm pert}}{p_t} = 
  \frac{\as}{\pi} \, L_i \,  \ln R + \order{\as} \, ,
\eeq
in a fixed-coupling approximation, with corrections which are non-singular as 
$R \to 0$. The main feature of \eq{eq:deltapt_pert_smallR_res} is the logarithmic
dependence on the radius $R$, which may spoil the convergence of the perturbative result at small $R$. For a complete description of the perturbative result at the smallest $R$ values, one might need to resum this logarithmic enhancement to all orders, an analysis which we postpone to future work. 

For completeness, we also compute the $R$-independent shift in transverse
momentum between the stable-cone-type and $k_t$ (or Cambridge) algorithms.
It is defined by
\beq
  \label{eq:dkt-cone}
  \frac{\langle \delta p_t^{\cone} \rangle_{\rm pert} - 
          \langle \delta p_t^{k_t} \rangle_{\rm pert}}{p_t} = 
          \frac{\as}{\pi} \, K_i 
\eeq
where again $i$ labels the parton species and we find
\beqa
  \label{eq:Kq}
  K_q & = & \int d z \min \big(z, 1 - z \big) \ln \big( f_\cone (z) \big) 
  P_{q q} (z) \nonumber \\ & = & \left(- \frac{15}{16} + \frac{9}{8} \ln
  2 + \ln^2{2} \right) \CF \simeq 0.323 \CF \, ,
\eeqa
for quarks, while for gluons one has
\beqa
  \label{eq:Kg}
  K_g & = & \int d z \min(z, 1 - z) \ln \big( f_\cone (z) \big) \left(
  \frac{1}{2} P_{g g} (z) + n_f P_{q g} \right) \\ & = &
  \left(- \frac{1321}{1152} + \frac{133}{96} \ln 2 + \ln^2 2 \right) \CA
  + \left(\frac{241}{576} - \frac{25}{48} \ln 2 \right) \nf \TR
  \simeq 0.294 \CA + 0.057 \nf \TR  \, . \nonumber
\eeqa
Numerically, one notes that the $K_i \sim 0.3 L_i $. This is the cause of the 
feature (originally observed in \cite{ESkt}) that,
perturbatively, $k_t$ and cone algorithms behave similarly when $\ln
R_{k_t} \simeq \, 0.3 + \ln R_\cone$, or equivalently $R_{k_t} \simeq
1.35 \, R_{\cone}$.

A similar analysis can be carried out for other observables as well. One can for example consider the single-inclusive jet $p_t$ distribution. Assuming a spectrum that falls off as $1/p_t^n$ at Born level, one finds at NLO a $\ln R$ correction
(in this case multiplying the Born level distribution $\propto 1/p_t^n$) which
is still of the form of \eqref{eq:deltapt_pert_smallR_res}, but with new 
coefficients $L_q (n)$ and $L_g (n)$.  Defining as usual
\beq
 \psi(n) = \frac{d}{dn} \ln \Gamma(n) = - \gamma_E + \sum_{p = 1}^{n - 1} 
 \frac{1}{p} \,, 
 \label{psin}
\eeq
one finds
\beqa
  \label{eq:Lqinc}
  L_q (n) & = & \int_0^1 d z \left( z^{n - 1} + (1 - z)^{n - 1} - 1 \right)
  P_{q q} (z) \nonumber \\ & = & - \, \CF \left(2 \psi(n - 1) + 
  2 \gamma_E + \left( \frac{3}{n}
  - \frac{3}{2} \right) \right)~,
\eeqa
as well as
\beqa
  \label{eq:Lginc}
  L_g (n) & = & \int d z \left( z^{n - 1} + (1 - z)^{n - 1} - 1\right) \left(
  \frac{1}{2} P_{g g} (z) + n_f P_{q g} \right) \nonumber \\
  & = & - \, \CA \left(2 \psi(n - 1) + 2 \gamma_E + \frac{4}{n} - \frac{2}{n+1}
  + \frac{2}{n + 2} - \frac{11}{6} \right) \\ &&  +  \, \, 
  2 \nf \TR \left(\frac{2}{n + 2}
  - \frac{2}{n + 1} + \frac{1}{n} - \frac{1}{3} \right)~. \nonumber
\eeqa
This pattern is general: for a generic jet observable, the logarithmic behaviour 
in $R$ is dictated by collinear dynamics, and can be computed at NLO (and
beyond) using only splitting function information rather than the full squared 
matrix element. In this regard, we note that Ref.~\cite{Ellis:2007ib} has 
discussed  the $R$ dependence of the average squared jet mass, observing
that to first order in $\as$ it behaves as $\as R^2 p_t^2$.

\section{An analytical estimate for hadronisation corrections}
\label{sec:analytical}

To be concrete, let us choose a definite process and observable. We consider 
single-inclusive jet production near partonic threshold in hadronic collisions, 
a process of current interest at the Tevatron and soon at the LHC. Our definition 
of threshold is that the scaled transverse momentum $x_t = 2 p_t/\sqrt{s}$
approaches its kinematical limit $x_t = 1$, so that all the available
collision energy is converted into the jet transverse momentum. We have in 
mind the rapidity-integrated distribution, which however is known to be well approximated near threshold by the value of the distribution at vanishing 
rapidity~\cite{deFlorian:2007fv}. We work then in the partonic centre-of-mass 
frame, and place the trigger jet at vanishing rapidity, with no loss of generality.
At Born level we have the kinematics
\beqa
 \label{eq:kin}
 p_1 & = & \frac{\sqrt{s}}{2} (1, 0, 0, 1) \\
 p_2 & = &  \frac{\sqrt{s}}{2} (1, 0, 0, -1)\nonumber \\
 p_3 = p_j & = &  p_t (1, 1, 0, 0) \nonumber \\
 p_4 = p_r & = &  p_t(1, -1, 0, 0) \nonumber
\eeqa
where $p_j$ is the four-momentum of the trigger jet, $p_t$ its transverse momentum and $p_r$  the four-momentum of the recoil jet. At Born level 
both jets correspond to massless particles and it follows that
\beq
 p_t = \frac{\sqrt{s}}{2}~.
 \label{ptthr}
\eeq

Consider now the change in $p_t$ induced by a soft gluon emission. 
We must separately consider two alternative scenarios: either the soft
emission is recombined by the jet algorithm with the hard parton associated 
with the trigger jet, or it is left unrecombined, so that the measured jet remains massless at this order. The recombination scheme we choose is the one commonly used for jet studies at the Tevatron, the $E$ scheme, where the four-momentum of the jet is obtained by adding the four-momenta of the constituent partons.

In the case where recombination happens the final state kinematics becomes
\beqa
 \label{recomb}
 p_j & = &  (\sqrt{p_t^2 + M_j^2}, p_t, 0, 0) \\
 p_r & = &  (p_t, - p_t, 0, 0)~, \nonumber
\eeqa
where the trigger jet now has a mass $M_j^2$, while the recoil jet is still a 
massless hard parton. Using energy conservation one has then
\beq
 p_t + \sqrt{p_t^2 + M_j^2} = \sqrt{s}~.
 \label{recomb2}
\eeq
Expanding to first order in the mass $M_j^2$ (since the gluon we recombine
with the jet has been assumed to be soft) we obtain
\beq
 p_t = \frac{\sqrt{s}}{2} \left(1 - \frac{M_j^2}{s} \right)~.
 \label{recomb3}
\eeq
The change in $p_t$ from its Born value is given by 
\beq
 \delta p_t^{+} (k) = - \frac{M_j^2}{2 \sqrt{s}} = 
 - \frac{p_3 \cdot k}{\sqrt{s}}~,
 \label{recomb4}
\eeq
where the $+$ superfix denotes the case when the gluon is recombined with 
the jet, and we wrote the jet mass as $M_j^2 = (p_3 + k)^2 = 2 p_3 \cdot k$,
with $p_3$ the four-momentum of the hard massless parton initiating the jet, 
and $k$ that of the soft gluon.

When the  gluon is not recombined, the recoil system is massive and one 
similarly finds
\beq
 \delta p_t^{-} (k) = - \frac{M_r^2}{2 \sqrt{s}} = 
 - \frac{p_4 \cdot k}{\sqrt{s}}~,
 \label{unrecomb}
\eeq
where $p_4$ denotes the four-momentum of the massless hard parton recoiling
against the trigger jet. Parametrising the soft gluon four-momentum as
\beq
 k^\mu  = k_t \left(\cosh \eta, \cos \phi, \sin \phi, \sinh \eta \right)~,
 \label{softparam}
\eeq
where $k_t$, $\eta$ and $\phi$ are respectively the transverse momentum, rapidity and azimuth defined with respect to the beam direction, it is easy to 
see that to first order in the small quantity $k_t$ one has 
\beq
 \label{delpt}
 \delta p_t^{\pm} (k) = - \frac{k_t}{2} \left(\cosh \eta \mp 
 \cos \phi \right)~.
\eeq
To obtain the average change in jet $p_t$ due to soft emissions, one has to
multiply the result in~\eq{delpt}, for the change in $p_t$ due to a soft emission, times the probability of emitting the soft gluon, and subsequently integrate over
phase space. In the eikonal approximation, appropriate to soft emissions, the
probability (squared matrix element) for the emission of a gluon with momentum
$k$ from an ensemble of hard partons with momenta $p_i$ may be expressed 
as a sum over contributions from all possible colour dipoles
\beq
 \label{eq:dipsum}
 \left| {\mathcal{M}} \right|^2 = \left| {\cal M}_0 \right|^2 
 \sum_{(ij)} C_{ij} \, W_{ij}(k)~,
\eeq 
where the sum runs over all distinct pairs $(ij)$ of hard partons, or equivalently, as stated before, over all dipoles. The quantity $\left| {\cal M}_0 \right|^2$ is the squared matrix element for the hard scattering, which in our case has to be computed for each separate partonic subprocess contributing to the jet distribution, and contains the dependence on parton distribution functions. 
The contribution of each dipole $W_{ij}$ is weighted by the colour factor 
$C_{i j} = - 2 \left(\bf{T_i} . \bf{T_j} \right)$, where $\bf{T_{i, j}}$ are generators of SU(3) corresponding to the colour charges of partons $i$ and $j$, 
while the kinematic factor $W_{i j} (k)$ is explicitly given by the classical 
antenna function
\beq
 W_{ij} (k) = \frac{\alpha_s \left( \kappa_{t, i j} \right)}{2 \pi}
 \frac{p_i \cdot p_j}{(p_i \cdot k)(p_j \cdot k)}
 \label{eikant}
\eeq
where $\alpha_s$ is defined in the bremsstrahlung scheme \cite{Catani:1990rr},
and its argument is the invariant quantity $\kappa_{t, i j}^2 = 2 (p_i \cdot k)(p_j
\cdot k)/(p_i \cdot p_j)$, which is just the transverse momentum with respect
to  the dipole axis, in the dipole rest frame. Assembling these results one can 
write an expression for the average shift in the jet transverse momentum when a soft gluon is emitted in a selected partonic channel. These shifts must then be recombined with the proper weights when the full cross section is built by 
summing over the various hard scattering processes contributing to a given distribution. In any selected channel we write
\beq
 \label{eq:dipint}
 \langle \delta p_t \rangle = \sum_{(ij)} C_{ij} 
 \int d k_t  \, k_t \, d \eta \, \frac{d \phi}{2 \pi} \, \frac{\alpha_s 
 \left( \kappa_{t, i j} \right)}{2 \pi} \, \frac{(p_i \cdot p_j)}{(p_i \cdot k)(p_j  
 \cdot k)} 
 \, \delta p_t(k)~,
\eeq  
where we integrated over the soft gluon phase space and we defined
\beq
 \delta p_t (k) = \delta p_t^{+} (k) \, \Theta_{\mathrm{in}} + 
 \delta p_t^{-} (k) \, \Theta_{\mathrm{out}}~,
 \label{thetas}
\eeq
with $\Theta_{\mathrm{in}}$ being unity if the gluon is inside the jet and zero
otherwise, and conversely for $\Theta_{\mathrm{out}}$.  To extract the part
of the soft contribution that one can associate with non-perturbative hadronisation
effects, $\langle \delta p_t \rangle_{\mathrm{h}}$,  one simply considers
the region $\kappa_{t, i j} < \mu_I $ in \eq{eq:dipint}, with $\mu_I$ an infrared
factorisation scale, to be chosen so that above $\mu_I$ one may safely use
perturbation theory. The coupling $\alpha_s$, which is perturbatively divergent
in this region, is now to be replaced by a universal, non-perturbatively defined,
finite quantity, whose moments at low energy are expected to be observable. 
In addition, one has to remove the contribution to \eq{eq:dipint} that would be
included in fixed-order perturbative contributions, so that one is left with a pure
hadronisation piece, which can subsequently be combined with fixed order
perturbation theory without double counting. In \eq{eq:dipint} we replace then 
the coupling $\alpha_s (\kappa_t)$ with $\delta \alpha_s(\kappa_t) = 
\alpha_s(\kappa_t) - \alpha_s^{\mathrm{PT}}(\kappa_t)$, where 
$\alpha_s^{\mathrm{PT}}(\kappa_t)$ is the standard perturbative coupling, 
which can be expanded in powers of $\alpha_s(p_t)$ to the desired order of
perturbation theory. We note that the standard choice for $\mu_I$ in the case 
of LEP event-shape studies was $2$ GeV, but it should be emphasised that 
the sensitivity to the choice of this scale is only $\mathcal{O} \left( 
\alpha_s^{n+1} \right)$, if one correctly combines the non-perturbative result 
with perturbative corrections evaluated to $\mathcal{O} \left(\alpha_s^n
\right)$~\cite{Dasgupta:2003iq}.

Let us continue by writing, purely for the sake of calculational convenience,
\beq
 \label{div}
 \delta p_t (k) = \delta p_t^{-} (k) + \left( \delta p_t^{+} (k) - 
 \delta p_t^{-} (k) \right) \, \Theta_{\mathrm{in}}~,
\eeq
where we used the fact that $\Theta_{\mathrm{out}} + \Theta_{\mathrm{in}}
= 1$. In this way, we have divided the integral in \eq{eq:dipint} into a `global' 
term, involving an integral over all of phase space of the unrecombined gluon
contribution $\delta p_t^{-}$, and a term involving an integral over the interior 
of the jet region. We demonstrate in the Appendix that the global term does not
produce a leading power correction, so that the complete leading 
contribution arises from the term involving the integral over the jet region.

We note here that at the level of the single-gluon calculation all the jet 
algorithms function in an identical manner, so that a soft gluon is recombined 
with a hard parton (and $\Theta_{\mathrm{in}} = 1$) if $\delta \eta^2 +
\delta \phi^2 < R^2$. In the present case, since we have fixed the trigger 
jet at $\eta = \phi = 0$, the gluon is in the jet for $\eta^2 + \phi^2 < R^2$.
As noted above, the only relevant contribution to the present accuracy arises 
from the second term in the sum on the RHS of \eq{div}. As a consequence, 
we only need to evaluate integrals of the form
\beq
 \label{eq:dipint2}
 \langle \delta p_t \rangle_{\mathrm{h}} = \sum_{i,j} 
 \, C_{ij} \hspace{-1mm} \int d k_t \, k_t \, d\eta \, \frac{d\phi}{2 \pi} \, 
 \frac{\delta\alpha_s \left(\kappa_{t, i j} \right)}{2\pi} \frac{(p_i 
 \cdot p_j)}{(p_i \cdot k)(p_j \cdot k)} \left( \delta p_t^+ - \delta p_t^- 
 \right ) \Theta_{\mathrm{in}} \, \theta \left (\mu_I - \kappa_{t, i j}\right) \, ,
\eeq 
for each dipole, recombined with the colour factors appropriate 
to the parton channel being considered.

\subsection{Dipoles involving the trigger jet and incoming partons}
\label{ssec:1j}

Consider first the dipole formed by one of the incoming partons (say the one 
with four momentum $p_1$) and the outgoing hard parton corresponding to 
the trigger jet, which we denote with the label $(1j)$. The transverse 
momentum with respect to this dipole is given by 
\beq
 \label{eq:katie}
 \kappa^2_{t, 1 j} \, = \, 2 \, \frac{(p_1 \cdot k)(p_j \cdot k)}{(p_1 
 \cdot p_j)} \, = \, 2 \, k_t^2  \, \left(\cosh \eta - \cos \phi 
 \right) \, .
\eeq 
Using $\delta p_t^{+} (k) - \delta p_t^{-} (k) = k_t \cos \phi$, and changing
variables from $k_t$ to $\kappa_{t, 1 j}$ (since the factorisation 
scale $\mu_I$ sets a limit on the range of $\kappa_{t, 1 j}$, not of $k_t$), 
we obtain
\beq
 \langle \delta p_t \rangle^{(1 j)}_{\mathrm{h}} =  
 \, C_{1j} \, \frac{1}{2} \, {\cal A} (\mu_I)  \int_{\eta^2 + \phi^2 < R^2} 
 d \eta \, \frac{d \phi}{2 \pi} \, {\rm e}^{3 \eta/2} \, \frac{\cos \phi}{\sqrt{2} 
 \left(\cosh \eta - \cos \phi \right)^{\frac{3}{2}}}~,
 \label{fin1J}
\eeq
where we defined the first moment of the non-perturbative coupling $\delta \alpha_s$ below the factorisation scale as
\beq
 {\cal A}(\mu_I) = \frac{1}{\pi} \int_0^{\mu_I} d \kappa_t \, \delta\alpha_s 
 \left( \kappa_t \right) \, .
 \label{cala}
\eeq
In order to make an explicit connection with event shape 
studies~\cite{Dasgupta:2003iq}, we can rewrite \eq{cala} in more detail as 
\beq
 {\cal A}(\mu_I) = \frac{1}{\pi} \, \mu_I \left [ \alpha_0 \left( \mu_I 
 \right) - \alpha_s (p_t) - \frac{\beta_0}{2 \pi} \left(\ln \frac{p_t}{\mu_I} +  
 \frac{K}{\beta_0} + 1 \right) \alpha_s^2 (p_t) \right] \, ,
 \label{likeevsh}
\eeq
where $\alpha_0 \equiv (1/\mu_I) \int_0^{\mu_I} \alpha_s(k_t) d k_t$ is
the average coupling over the infrared region, familiar from event shape studies,
and we have carried out the subtraction of the perturbative coupling, $\alpha_s
(p_t)$, to two-loop accuracy in the $\MSbar$ scheme, where 
$K = C_A \left( \frac{67}{18} - \frac{\pi^2}{6} \right) - \frac{5}{9}
n_f$. Note that similar expressions in Ref.~\cite{Dasgupta:2003iq} are rescaled
by the Milan factor, accounting for gluon decays; this rescaling will be discussed
below, in~\secn{ssec:leading}. 

The integral over the soft gluon direction can be evaluated by choosing polar
coordinates in the $\eta-\phi$ plane and expanding in powers of the radial
variable.  Discarding the spurious collinear divergence arising when the gluon 
is emitted along the outgoing leg, which cancels against an identical one in 
the `global' term, as shown in the Appendix, one finds
\beq
 \langle \delta p_t \rangle^{(1j)}_{\mathrm{h}}  = 
 \, C_{1j}  \, {\cal A} (\mu_I) 
 \left(- \frac{1}{R} + \frac{5}{16} R - \frac{23}{3072} R^3 - 
 \frac{95}{147456}R^5 + \mathcal{O} \left( R^7 \right) \right)~.
 \label{finfin1J}
\eeq
By symmetry, an identical result is obtained for the $2j$ dipole, formed by 
the trigger hard parton and the other incoming parton with momentum $p_2$.

\subsection{Dipole involving the trigger and recoil jets}
\label{ssec:jr}

In this case the transverse momentum of the soft gluon with respect to the 
dipole, which we label with $(jr)$, is given by
\beq
 \label{eq:katjr}
 \kappa^2_{t, j r} \, = \, k_t^2 \, e^{-\eta} \, \left(\cosh^2 \eta - 
 \cos^2 \phi \right)~,
\eeq 
which leads to the integral
\beq
 \langle \delta p_t \rangle^{(jr)}_{\mathrm{h}} =  \,
 C_{jr} \,  {\cal A} (\mu_I) \int_{\eta^2 + \phi^2 < R^2}
 d \eta \, \frac{d \phi}{2 \pi} \, \frac{\cos \phi}{
 \left(\cosh^2 \eta - \cos^2 \phi \right)^{3/2}}~.
 \label{deljr}
\eeq
This can be evaluated as before, using polar coordinates and expanding in 
powers of $R$. One finds
\beq
 \langle \delta p_t \rangle^{(jr)}_{\mathrm{h}}   =  \,
 C_{jr} \,  \mathcal{A}(\mu_I) \, \left(-\frac{1}{R} - 
 \frac{1}{4} R + \frac{1}{192} R^3 - \frac{5}{2304}R^5 + 
 \mathcal{O} \left( R^7\right) \right)~.
 \label{finJR}
\eeq

\subsection{Incoming dipole}
\label{ssec:12}

The dipole involving the two incoming partons, labelled as $(12)$, is the simplest
to compute, since in this case $\kappa_{t, 12} = k_t$. For this dipole the
integration over the interior of the jet does not produce a $1/R$ correction, 
since there is no collinear enhancement for radiation emitted by the incoming
partons as the jet becomes narrow. We expect radiation from this dipole to 
behave essentially like an underlying event contribution, and this is in fact what 
we find. The relevant integral is 
\beqa
 \langle \delta p_t \rangle ^{(12)}_{\mathrm{h}} & = & \,
 C_{12} \, {\cal A} (\mu_I)  \,
 \int_{\eta^2 + \phi^2 < R^2} d \eta \, \, 
 \frac{d \phi}{2 \pi} \, \cos \phi \nonumber \\
 & = & \, C_{12} \, {\cal A} (\mu_I)  \, R \, J_1 (R) \nonumber \\ 
 & = & \, C_{12} \, {\cal A} (\mu_I)  \, \left( \frac{1}{2} R^2 - 
 \frac{1}{16} R^4 + \frac{1}{384} R^6 + \mathcal{O} \left( R^8 \right) 
 \right)~,
 \label{fin12}
\eeqa
where $J_1$ is the Bessel function of the first kind.

\subsection{Dipoles involving incoming partons and the recoil jet}
\label{ssec:1r}

As these dipoles involve radiation which is not associated with the trigger jet,
we expect, and verify, that they do not generate any $1/R$ enhancement.
Considering for example the $(1r)$ dipole, the relevant transverse momentum
is
\beq
 \label{eq:kat1r}
 \kappa^2_{t, 1 r} \, = 2 \, k_t^2 \, e^{-\eta} \, \left(\cosh \eta + 
 \cos \phi \right)~,
\eeq 
which leads to the integral
\beq
 \langle \delta p_t \rangle^{(1 r)}_{\mathrm{h}} = \, 
 C_{1 r} \, \frac{1}{2} {\cal A} (\mu_I) \, \int_{\eta^2 + \phi^2 < R^2} 
 d \eta \,  \frac{d \phi}{2\pi} \, \frac{e^{\frac{3\eta}{2}} \cos 
 \phi}{\sqrt{2} \left(\cosh \eta + \cos\phi \right )^\frac{3}{2}}~.
\eeq
This gives
\beq
 \langle \delta p_t \rangle ^{1 r}_{\mathrm{h}}  = \,
 C_{1 r}  \, {\mathcal{A}}(\mu_I)  \left( \frac{1}{16} R^2 + 
 \frac{5}{512} R^4 - \frac{95}{49152} R^6 + 
 \mathcal{O} \left( R^8 \right) \right)~.
 \label{fin1R}
\eeq
Clearly, the $(2 r)$ dipole gives an identical contribution.

\subsection{Leading power correction}
\label{ssec:leading}

Having computed the results for each individual dipole, we can now
perform the sum in \eq{eq:dipsum} to obtain our estimate for the shift in the 
jet $p_t$ due to hadronisation effects. We concentrate on the leading behaviour 
at small $R$, including terms up to ${\cal O} (R)$, since at large values of $R$ 
we expect the hadronisation component to be dominated by the underlying event
contribution, which behaves like $R^2$. Clearly, each parton channel will have
a different $p_t$ shift, because of the different colour weights in \eq{eq:dipsum}.
Here we choose as an example, to illustrate our point, the scattering of 
non-identical quarks,
\beq
 q(p_1) + q'(p_2) \to q(p_3) + q'(p_4)~.
 \label{qqp}
\eeq
The calculation of the colour weight of the various dipoles in \eq{eq:dipsum} is
a textbook exercise~\cite{Ellis:1991qj}. The result is 
\beq
 \left| {\cal M}^{qq' \to qq'g}\right|^2 = \left| {\cal M}_0^{qq' \to qq'} \right|^2
 \left( 2 C_F \left(W_{14} + W_{23} \right) + \frac{1}{N_c} 
 \left[W_{12}; \, W_{34}\right] \right)~,
 \label{diptog}
\eeq
where, in a notation similar to~\cite{Ellis:1991qj}, we have defined
\beq
 \left[W_{12}; \, W_{34} \right] = 2 W_{12} + 2 W_{34} - W_{13} - W_{14}
 - W_{23} - W_{24}~,
 \label{dipfin}
\eeq 
which is a collinear finite combination of dipoles contributing at subleading order
in the number of colours $N_c$. Since we are interested in just the $1/R$ and $R$ 
terms, we need only consider dipoles that involve the hard parton responsible for the trigger jet, which we identify with the quark carrying momentum $p_3$.
Adding together the contributions from the relevant dipoles one easily finds
\beq
 \langle \delta p_t \rangle_{\mathrm{h}}^{qq' \to qq'} = 
 {\cal A} (\mu_I) \left [- \frac{2}{R}
 C_F + \frac{1}{8} R \left( 5 C_F - \frac{9}{N_c} \right) 
 + {\cal O} \left( R^2 \right) \right]~.
 \label{finpt}
\eeq
We note several features of \eq{finpt}, most of which are common to the
other parton channels as well. The most striking aspect is clearly the singular
dependence on the jet radius $R$, arising from all dipoles involving the trigger
jet. This singular behaviour is a reflection of the fact that, as one makes the jet
narrower, one increases the $p_t$ loss due to hadronisation, which also explains 
the negative sign of the leading term. The leading behaviour as $R \to 0$
arises from the collinear singularity of the matrix element, which is also 
responsible for the $\ln R$ enhancement of the perturbative result
described in Sect.~\ref{sec:pertR}. For this reason, one expects the $1/R$ 
behaviour to be accompanied by the colour charge of the parton initiating the 
jet, which in the above case is $C_F$, and this expectation is indeed 
confirmed when one combines the dipoles. When the trigger jet originates 
from a gluon, as it would be for instance in the $gg \to gg$ channel, the 
leading $1/R$ behaviour of the hadronisation correction can thus be simply
obtained with the replacement $C_F \to C_A$. 

Another point to note about the $1/R$ term in \eq{finpt} is the fact that its
coefficient,  $2 \, C_F {\cal A} (\mu_I)$, is one half of the coefficient of the 
$1/Q$ correction to the thrust variable in $e^+ e^-$ annihilation, as computed
in \cite{Dokshitzer:1995zt}. In order to match more recent evaluations
\cite{Dokshitzer:1997iz,Dokshitzer:1998pt,Dasgupta:1998xt,Dasgupta:1999mb,Smye:2001gq}, this result for the thrust must actually be rescaled 
by a factor $\frac{2}{\pi} M$, where $M$ is the Milan factor. Specifically,
the factor $\frac{2}{\pi}$ accounts for the use of a fully dispersive coupling 
(one expressed in terms of the gluon virtuality rather than its transverse
momentum), while the Milan factor $M$ accounts for the non-inclusiveness 
of final-state observables, \ie the fact that a virtual gluon inevitably splits 
into massless objects, but the observable may change its value unless all
of the gluon decay products are emitted into the appropriate region of
phase space. For observables that are linear in multiple soft-gluon momenta, 
the Milan factor is independent of all other details of the observable, and
numerically one finds $M \simeq 1.49$ (for $n_f = 3$ in the non-perturbative
region), or equivalently $\frac{2}{\pi} M \simeq 0.95$, which is a modest 
overall correction. We note, however, that jet transverse momenta are in 
general \emph{not} linear in soft-gluon momenta, because of the non-trivial 
way in which soft-particles recombine with each other and into jets. The one 
known exception in this respect, among infrared and collinear safe jet algorithms, 
is the anti-$k_t$ algorithm\footnote{More generally, all algorithms with
 $p<0$ in \eq{eq:seq-rec}.}, introduced and discussed at length in
\cite{antikt}.  Thus, while at present we can tentatively associate the
size of our $1/R$ effect with the thrust calculation, this implicitly
assumes that an analysis similar to that performed in
Refs.~\cite{Dokshitzer:1997iz,Dokshitzer:1998pt,Dasgupta:1999mb} will
yield results that do not differ significantly from the value computed
for $M$ in the case of linear observables.

We finally observe that the coefficient of the term linear in $R$ in \eq{finpt}
is numerically less than $20 \%$ of the coefficient of the $1/R$ term. This 
means that the $1/R$ behaviour should be a good approximation to the full 
result over a wide range of values of $R$,  up to $R \sim 1$, a fact which 
will be useful to construct simple approximate formulae in 
\secn{sec:exp-considerations}.

\subsection{Mass distribution}
\label{ssec:mass}

In order to illustrate the generality of the method, let us briefly summarise the
results one obtains when performing the same computation with a different
observable. We now compute the hadronisation correction to the jet mass, 
$\delta M_j^2$, using the framework described above. Working again near 
threshold, the variation in the jet mass due to the emission of a single soft 
gluon is given by
\beq
 \delta M_j^2 \left(k_t, \eta, \phi \right) = 2 p_j \cdot k = 
 k_t \sqrt{s} \left( \cosh \eta - \cos \phi \right)~,
 \label{nuin}
\eeq
provided the jet algorithm recombines the gluon with the jet. Clearly, there is no correction if the gluon is not recombined. As the singular $1/R$ behaviour of the hadronisation correction to the transverse momentum of the jet was due to
\emph{unrecombined} gluons becoming increasingly collinear to the originating
hard parton when $R\to 0$, there should be no corresponding singular piece 
for the jet mass. We find in fact that the leading power corrections vanish 
at least as $R$ for all dipoles. Specifically
\beqa
 \label{jmdip} 
 \delta M^2_{j, 1 2} & = & \, C_{12} \, 
 {\mathcal{A}} (\mu_I) \, \frac{\sqrt{s}}{2} \, \left( \frac{1}{4} R^4 +
   \frac{1}{4608} R^8 + {\cal O} \left( R^{12} \right) \right) \, ,
   \nonumber \\
 \delta M^2_{j, 1 j} & = & \, C_{1j} \,  {\mathcal{A}}
   (\mu_I) \, \frac{\sqrt{s}}{2} \, \left( R + \frac{3}{16} R^3 +
   \frac{125}{9216} R^5 + \frac{7}{16384} R^7 + {\cal O} \left( R^9
   \right) \right) \, ,
   \nonumber  \\
 \delta M^2_{j, j r} & = & \, C_{jr} \,  {\mathcal{A}}
   (\mu_I) \, \frac{\sqrt{s}}{2} \, \left( R + \frac{5}{576} R^5 + {\cal O}
   \left( R^9 \right) \right) \, ,
   \\
 \delta M^2_{j, 1 r} & = & \, C_{1r} \,  {\mathcal{A}}
   (\mu_I) \frac{\sqrt{s}}{2} \, \left( \frac{1}{32} R^4 + \frac{3}{256} R^6 +
   \frac{169}{589824} R^8 + {\cal O} \left( R^{10} \right) \right) \, .
   \nonumber   
\eeqa
We observe that the small radius behaviour of the leading power correction to 
the jet mass distribution is softened with respect to the $p_t$ distribution by 
two powers of $R$. This is a consequence of the fact that gluons emitted inside small cones make small contributions to the jet mass, just as for the $p_t$ distribution. Gluons which are not recombined with the jet, on the other hand, in
this case make no contribution, so that the singular behaviour at small $R$ is
absent.  Note however that the contribution of the underlying event to the jet
mass is ${\cal O} (R^4)$, so that again the hadronisation component dominates 
by three powers of $R$, for $R < 1$.

\section{Comparison with Monte Carlo results}
\label{sec:montecarlo}

Two simple results emerge from the analytic approach of 
\secn{sec:analytical}, concerning our expectations for the 
non-perturbative modification of a jet transverse momentum. First, 
non-perturbative radiation associated with emission from the jet's 
originating parton, in a selected production channel, should modify 
the jet $p_t$ as
\beq
 \label{eq:summary_hadronization}
 \langle \delta p_t \rangle_h 
 = - \, C_i \, \frac{2}{R} \, {\cal A} (\mu_I) + \order{R} \, ,
\eeq
where $C_i$ is the colour charge appropriate to the parton originating
the jet\footnote{As noted in the Introduction, this form of $R$ dependence 
 for the leading power correction was already suggested 
 in~\cite{Korchemsky:1994is} and its full form could probably have been 
 deduced from~\cite{Seymour:1997kj}. It seems, however, from discussions 
 of hadronisation in the literature~\cite{Mangano:1999sz,Campbell:2006wx}, 
 that the community was not aware of the implications of these results.}.
Recall that, in the single soft gluon approximation the scale ${\cal A} (\mu_I)$ 
is related to the scale appearing in analytical studies of hadronisation in
$e^+e^-$ and DIS collisions. A first test of our approach would then be to fit 
data for jet distributions using expressions like \eq{eq:summary_hadronization},
with a numerical value for ${\cal A} (\mu_I)$ borrowed from event shape 
studies, including the appropriate Milan factor, as discussed in 
Sect.~\ref{ssec:leading}. We must then expect some degree of loss of
universality across different algorithms and observables, which introduce
different nonlinear multi-gluon effects. For our illustrative purposes, however, 
it is sufficient to be aware that $2 C_F \, {\cal A} (2 {\rm GeV}) 
\simeq 0.5 \GeV$, corresponding to the amount of non-perturbative 
transverse momentum radiated per unit rapidity with respect to a 
$q \bar q$ dipole, while a $gg$ dipole must be reweighted by a factor
$C_A/C_F = 9/4$. 

Our second result is that corrections associated with the dipole of incoming 
partons, and in fact more generally with any dipole not involving the trigger 
jet, give a contribution that scales as $R^2$, \ie in proportion to the jet area.
These corrections effectively mimic underlying event corrections, and are indistinguishable from them. Borrowing our result in Sect.~\ref{ssec:12}, 
and denoting the scale of transverse momentum emission per unit rapidity 
along the beam direction with $\Lambda_{UE}$, we expect the jet
transverse momentum to be modified as\footnote{The terms of higher order 
in $R$ are specific to our chosen recombination scheme, the $E$-scheme or
four-vector recombination.}
\beq
 \label{eq:summary_pp}
 \langle \delta p_t \rangle_{\text{UE}} 
 = \Lambda_{UE} \, R \,  J_1(R) = \frac{\Lambda_{UE}}{2} \left(R^2 -
 R^4/8 + \order{R^6}\right) \, .
\eeq
The scale $\Lambda_{UE}$, as the notation suggests, cannot be easily 
related to ${\cal A} (\mu_I)$, since it receives contributions from the 
underlying event, \ie the interactions between the proton remnants. The
functional dependence on $R$ through the combination $R J_1 (R)$ 
corresponds to the definition of `passive vector area' of a jet given in 
Ref.~\cite{Cacciari:2007fd,CCSArea}.

It is of interest to compare the two sets of contributions, 
\eq{eq:summary_hadronization} and \eq{eq:summary_pp}, to
what is observed in parton-shower Monte Carlo event generators. 
To do so, we begin by generating dijet events; after parton showering,
we identify jets with a chosen jet algorithm, and label the transverse 
momenta of the two hardest jets as $p_{t, \ps}^{(i)}$, $(i = 1,2)$;
we select events for which the hardest jet satisfies $55\GeV < p_{t, \ps}^{(1)} 
< 70 \GeV$; next, we allow either the hadronization stage to take place, 
or both hadronization and underlying event generation; finally, we rerun the 
jet finder. After hadronization the two jet transverse momenta will be changed,
and we denote them with $p_{t, \h}^{(i)}$; similarly, after generating also the
underlying event, the momenta will become $p_{t, \UE}^{(i)}$. At this point, 
we can define the hadronization contribution to the jet transverse momentum 
to be
\beq
 \label{eq:dpthadr-mc}
 \langle \delta p_t \rangle_{\h}  = \frac12 \, \left\langle
 p_{t, \h}^{(1)} - p_{t, \ps}^{(1)} + p_{t, \h}^{(2)} - p_{t, \ps}^{(2)} 
 \right\rangle~,
\eeq
while the underlying event contribution is 
\beq
 \label{eq:dpthadr-UE}
 \langle \delta p_t \rangle_{\UE}  = \frac12 \left\langle
 p_{t, \UE}^{(1)} - p_{t, \ps}^{(1)} + p_{t, \UE}^{(2)} - p_{t, \ps}^{(2)} 
 \right\rangle - \langle \delta p_t \rangle_\h \, .
\eeq
Note that with most modern underlying event (UE) models it is not possible 
to carry out the UE generation independently from hadronization: one must 
first determine $\langle \delta p_t \rangle_\h$ with the global UE switch 
turned off, and subsequently carry out a separate run with the switch turned 
on, in order to deduce $\langle \delta p_t \rangle_\UE$.

\begin{figure}
  \centering
  \includegraphics[angle=270,scale=0.6]{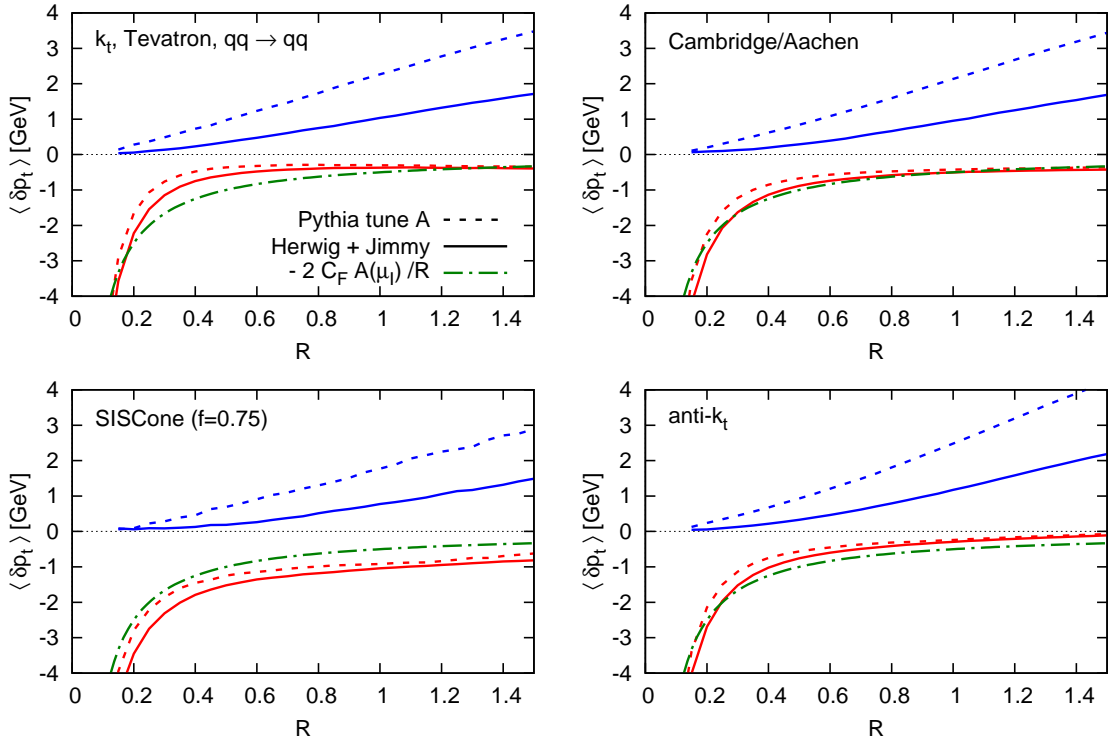}
  \caption{Modification of the $p_t$ of jets due to the underlying
    event (upper curves) and hadronization (lower curves), for $qq \to
    qq$ scattering at the Tevatron Run II ($p \bar p$,
    $\sqrt{s} = 1.96 \TeV$), comparing {\tt Pythia} 6.412~\cite{Sjostrand:2006za} 
    (tune A) and {\tt Herwig} 6.510~\cite{Corcella:2002jc} with 
    {\tt Jimmy} 4.3~\cite{Butterworth:1996zw}. 
    In the case of hadronization the Monte Carlo outputs are compared to
    the analytical result, \eq{eq:summary_hadronization}. Dijet events are 
    selected containing an underlying $qq \to qq$ scattering, and with the
    requirement that at parton shower level the hardest jet has $55
    \GeV < p_{t, \ps} < 70 \GeV$. The non-perturbative corrections 
    shown correspond to the average for the two hardest jets.}
  \label{fig:4stamp}
\end{figure}

Our results are shown in Fig.~\ref{fig:4stamp} for four different jet
algorithms, as a function of the jet radius $R$, for $qq \to qq$
interactions (summing over the flavors and antiflavours of the incoming
and outgoing quarks) in $p \bar p$ collisions, in Tevatron Run~II kinematics,
$\sqrt{s} = 1.96 \TeV$. The upper curves in each plot are for $\langle
\delta p_t \rangle_\UE$, while the lower ones show $\langle \delta
p_t \rangle_\h$. In each case we show results from 
{\tt Pythia}~6.4~\cite{Sjostrand:2006za}, (tune~A, using the 
default, `old' shower), and from 
{\tt Herwig}~6.5~\cite{Corcella:2002jc} with the {\tt Jimmy} UE 
model~\cite{Butterworth:1996zw} (with the Atlas
tune~\cite{Albrow:2006rt}). Additionally, for the hadronization 
results, we show the prediction based on the one-gluon emission approximation 
and universality assumption, \eq{eq:summary_hadronization}.

The four jet algorithms we use are all infrared and collinear (IRC)
safe, as they must be for any analysis beyond leading order.  Specifically,
three of the algorithms we use are sequential recombination algorithms; they operate by defining interparticle and beam-particle distance measures via
\beq
  \label{eq:seq-rec}
  d_{ij}^{(p)} \equiv \min \! \left( k_{t, i}^{2 p}, k_{t, j}^{2 p} \right) 
  \frac{\Delta y_{i j}^2 + \Delta \phi_{i j}^2}{R^2} \, , \quad \qquad 
  d_{i B}^{(p)} \equiv k_{t, i}^{2 p} \, ,
\eeq
and then recombining particles $i$ and $j$ if one of the $d_{i j}^{(p)}$ is 
smallest, or defining particle $i$ as a jet, and removing it from the list of 
particles, if $d_{i B}^{(p)}$ is smallest; the procedure is then iterated until no 
particles are left. The choice $p = 1$ corresponds to the well-known (inclusive)
$k_t$ algorithm~\cite{inclukt,ESkt}. Choosing $p = 0$ gives the Cambridge/Aachen 
algorithm~\cite{Caachen}, probably the simplest possible IRC-safe 
hadron-collider jet algorithm, corresponding to the successive recombination 
of particles that are closest on the $y-\phi$ cylinder, until all are separated by 
at least $R$. Its simplicity makes it our preferred choice in those figures 
where for brevity we consider only a single algorithm.
Our third choice, $p = - 1$, dubbed `anti-$k_t$' algorithm, is 
novel~\cite{antikt}, and has several interesting properties: for example, 
it behaves like a perfect cone algorithm, in that hard jets are nearly always 
circular, with radius $R$ on the $y-\phi$ cylinder; furthermore, it behaves linearly 
in soft gluon momenta, so that its hadronization corrections should have better
universality properties, as was the case for event shapes in $e^+ e^-$ annihilation.
As a fourth example, we use a seedless stable-cone algorithm, the SISCone
algorithm~\cite{Salam:2007xv}, with a Tevatron run~II type of split-merge 
procedure~\cite{Blazey}. It is similar to the midpoint algorithm~\cite{Blazey}
used at Tevatron, differing mainly in its seedless nature and the resulting fact 
that it is IRC safe at all orders.

A striking feature of Fig.~\ref{fig:4stamp} is that all four jet
algorithms have similar $R$ dependences: despite much debate about the
relative merits of different algorithms, they all behave identically
in the presence of a single soft gluon, and, as a consequence, there are
strong similarities in their non-perturbative behaviour.

The non-perturbative contribution that is generally acknowledged as
being best modelled is hadronization: indeed the Monte Carlo models agree 
well between each other, as is natural given the extensive tuning of the
hadronization-related parameters at LEP. The analytical prediction, 
\eq{eq:summary_hadronization}, reproduces both the shape and normalisation
relatively well, though the quality of agreement varies somewhat with the
algorithm: the $k_t$ (SISCone) algorithm, for example,  has slightly less (more)
negative corrections than predicted. These differences may be a reflection of 
the breakdown of the single gluon approximation, caused by non-linearity of the $k_t$, Cambridge/Aachen and SISCone algorithms. A full treatment would
involve the inclusion in \eq{eq:summary_hadronization} of the
non-trivial corrections associated with double-soft gluon emission,
\ie the (non-universal) Milan factor. Clearly, this would be of interest for 
further study.

\begin{figure}
  \centering
  \includegraphics[angle=-90,scale=0.6]{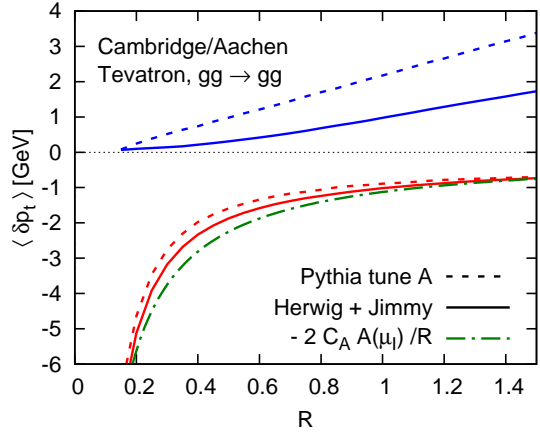}
  \caption{Similar to Fig.~\ref{fig:4stamp}, but for the $gg \to gg$
    underlying scattering channel, instead of $qq \to qq$. For brevity
    only the Cambridge/Aachen result is shown.}
  \label{fig:tev-gg}
\end{figure}

\begin{figure}
  \centering
  \includegraphics[angle=-90,scale=0.6]{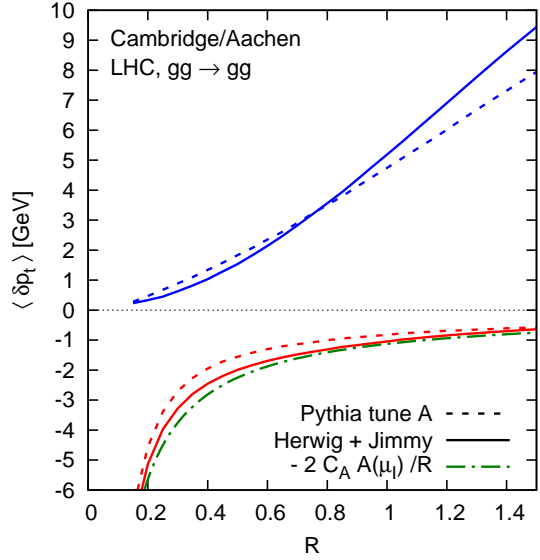}
  \caption{Similar to Fig.~\ref{fig:tev-gg}, but for the LHC ($pp$,
    $\sqrt{s}=14\TeV$) rather than the Tevatron. }
  \label{fig:lhc-gg}
\end{figure}

The dependence of the hadronization contribution on the jet colour
factor is visible in Fig.~\ref{fig:tev-gg}, which displays results for 
the $gg\to gg$ channel at Tevatron. We see that the hadronization 
contribution roughly doubles compared to $qq \to qq$ scattering, again 
in reasonable agreement with the analytical prediction. Fig.~\ref{fig:lhc-gg}, 
on the other hand, shows the independence of the hadronization contribution 
on collider energy, displaying results for $gg \to gg$ scattering at LHC, which 
are easily seen to be almost identical to the Tevatron results.

Underlying event contributions are much less well understood than
hadronization. Not only is no analytical prediction available for their
normalisation, but {\tt Pythia} tune~A and the {\tt Jimmy} Atlas tune, despite both being tuned to similar Tevatron data, give underlying event correction
to jet $p_t$'s that differ by a factor of two at the Tevatron, as is
visible in Figs.~\ref{fig:4stamp} and \ref{fig:tev-gg}. As can also be seen 
from these figures, the UE contribution is largely independent of the hard 
scattering channel ($q q \to q q$ versus\ $g g \to g g$). It does however 
depend strongly on the collider energy, as seen in Fig.~\ref{fig:lhc-gg}, 
which illustrates the huge size of the UE contribution predicted by Monte 
Carlo models at LHC. Interestingly, though {\tt Jimmy} and {\tt Pythia} differ 
for the Tevatron, they agree for LHC\footnote{We have also examined other 
{\tt Pythia} tunes (DW, DWT, S0, S0A). All are nearly identical at Tevatron 
energies, while, at LHC, tunes DW and S0A are similar to tune A, and tunes
S0 and DWT give results that are about $40-50\%$ higher.}.

\begin{figure}
  \centering
  \includegraphics[angle=-90,width=0.5\textwidth]{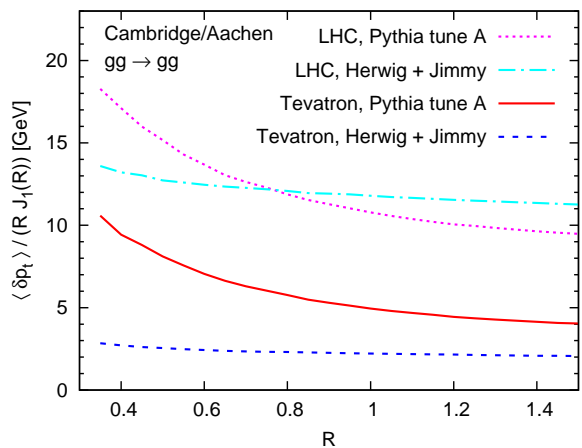}
  \caption{Plot of the UE contribution to the jet $p_t$, for Tevatron
    and LHC $gg\to gg$ events (selected as in Fig.~\ref{fig:4stamp}),
    rescaled by the factor $R J_1(R)$, corresponding to the passive 
    area~\cite{Cacciari:2007fd}, as calculated for a 1-particle jet.}
  \label{fig:scaled-UE}
\end{figure}

Let us now turn to the functional form of the UE corrections,
\eq{eq:summary_pp}, and the interpretation of its normalisation.
We consider $\langle \delta p_t \rangle_\UE$ divided by its predicted
$R$ dependence, given by $R J_1(R)$; this should give an $R$-independent 
result equal to $\Lambda_\UE$. The results are shown in 
Fig.~\ref{fig:scaled-UE}. Whereas for {\tt Jimmy} there is near perfect 
agreement with the scaling prediction, with {\tt Pythia} there is clear deviation 
from it towards small\footnote{Note however that we are unable to consider 
the smallest $R$ values, because the subtraction procedure 
in~\eq{eq:dpthadr-UE} makes it difficult to determine accurately
the small UE contribution relative to the large hadronization effect.} 
values of $R$, which can be interpreted as a correlation between underlying 
event activity and the properties of the hard scattering. Further 
investigation reveals that this is caused by strong colour reconnections 
in the {\tt Pythia} underlying event model and tunes, an effect which is 
enhanced for $g g \to g g$ scattering as compared to  $q q \to q q$ 
scattering
\footnote{This feature is present to the same degree in tunes DW, DWT
  based on the old shower, and in all cases in which the old shower is used, 
  it seems to be independent of the precise procedure used to extracts the UE
  contribution.  In tunes S0 and S0A, based on the new shower, it
  seems that the effect may be less strong, however it is difficult to
  make a firm statement because at small and moderate $R$ the
  extraction of $\langle \delta p_t\rangle_\UE$,
  eq.~(\ref{eq:dpthadr-UE}), is affected by non-trivial interplays
  between the parton shower and the UE multiple interactions.}. 
Note in any case that the physical impact of this correlation is limited, 
since the relative deviation from the expected $R J_1(R)$ form is significant 
only in the region where the absolute size of the UE contribution is modest.

Finally, let us comment on the physical scale associated with the
underlying event, $\Lambda_\UE$. At the Tevatron the value that
emerges from the models is $\Lambda_\UE(1.96\TeV) \simeq 2-4 \GeV$,
while at LHC it is $\Lambda_\UE(14\TeV) \simeq 10 \GeV$. This is an
order of magnitude larger than the corresponding scale for
hadronization of a quark ($2 C_F {\cal A} \simeq 0.5 \GeV$) or a gluon
($2 C_A {\cal A} \simeq 1.0 \GeV$). In all cases the scale relates to
the amount of radiation per unit rapidity with respect to the emitter.
The much larger scale associated with UE is not unexpected: one
can either interpret it in terms of multiple gluon-gluon interactions
between the proton remnants, or alternatively in terms of an effective
perturbative saturation scale, induced by small-$x$ dynamics; both
would lead to substantially higher values of the effective scale than expected
for normal hadronization.

Though not surprising, the large value of $\Lambda_\UE$ does have
implications for choices of $R$ in jet finding. Arguments based on
purely perturbative considerations (and on normal hadronization)
suggest that $R = 1$ is an optimal value. The non-perturbative
scale associated with the UE is so large, however, that for moderate 
$p_t$ jets it can easily be comparable to the perturbative contributions, 
proportional to $\as p_t$. 
This, for example, is the explanation for the poor behaviour 
of the $k_t$ algorithm in~\cite{D0kt} with the `recommended' $R = 1$ 
choice, while smaller $R$ values, used more recently for the $k_t$
algorithm in~\cite{CDFkt}, lead to more reasonable UE corrections. In
contrast, when examining very high $p_t$ jets, since the UE should be
$p_t$ independent, it will become advantageous to return to $R\sim 1$,
so as to minimise the size of perturbative corrections, which behave
roughly as $\sim \as p_t \ln R$.

\section{Experimental considerations}
\label{sec:exp-considerations}

There are a number of respects in which the results presented above
are relevant experimentally. On one hand, knowledge of the relative
size of various non-perturbative and perturbative contributions as a
function of $R$ can provide guidance in the choice of optimal values 
of $R$. On the other hand, experimental data can provide interesting
cross-checks of our understanding of the $R$ dependence, for example 
through direct measurements of how the transverse momentum of 
a jet depends on the parameters  of jet algorithm, and through studies
of quantities such as inclusive jet cross sections. In what follows, we will 
sketch three possible avenues of experimental investigation on these 
issues; they should be seen as examples of what could be done, which will 
have to be adapted to the chosen observable and  to the needs of the 
given experimental setup.

\subsection{Radius optimisation}
\label{sec:roptimization}

We have seen that there are three main sources of corrections to the
transverse momentum of a jet: perturbative radiation; hadronization, \ie
non-perturbative effects associated with the jet itself; and the
underlying event, \ie low-$p_t$ effects associated with
proton-remnant interactions. Each of these effects has a different
dependence on the jet $p_t$, on the colour factor associated with the
hard parton originating the jet, on the choice of $R$ in the jet finder,  and finally 
on the collider energy, as illustrated in table~\ref{tab:summary-of-effects}.

\begin{table}
  \centering
  \begin{tabular}{l|c|c|c|c|}
    & \multicolumn{4}{c|}{Dependence of jet $\langle \delta p_t \rangle$ on}\\
    & `partonic' $p_t$ & colour factor &  $R$ & $\sqrt{s}$\\\hline
    perturbative radiation & $\sim \as(p_t) \, p_t$ & $C_i$ & $\ln R +
    \order{1}$ & -- \\
    hadronization   & -- & $C_i$ & $-1/R + \order{R}$ &  -- \\
    underlying event  & -- & -- & $R^2 + \order{R^4}$ &  $s^{\omega}$ \\\hline
  \end{tabular}
  \caption{Summary of the main physical effects that contribute
    to the relation between the transverse momentum of a jet and that of a
    parton, together with their dependence on the properties of the parton, 
    the jet radius $R$ and collider centre of mass energy. Cases labelled 
    ``--'' do not have any dependence on the corresponding variable in a 
    leading approximation, but may develop anomalous-dimension type 
    dependences at higher orders.} 
  \label{tab:summary-of-effects}
\end{table}

Let us examine the implications of this understanding on the choice of
$R$ in various experimental contexts. There are two principal scenarios 
to be considered, corresponding to two rather different usages of jets.
One may use jets for the identification of underlying kinematic
structures, such as mass peaks for the top quark or other heavy particles, 
or in a search for hadronic decays of a hypothetical $Z'$.
Alternatively, one may wish to compare jet data (such as the inclusive
jet spectrum)
with high-order perturbative QCD calculations, and attempt to deduce
information about fundamentals of QCD or the electroweak theory. In
the first case, one seeks to extract the cleanest possible kinematic
structures, and therefore one should minimise both perturbative and
non-perturbative modifications of a jet $p_t$; in the second case,
one presumes that the perturbative loss is calculated with good
precision for typical ranges of $R$, and one wishes to minimise the 
two non-perturbative contributions, since they cannot be precisely computed 
from first principles.

\begin{figure}[ht]
  \centering\includegraphics[width=0.5\textwidth]{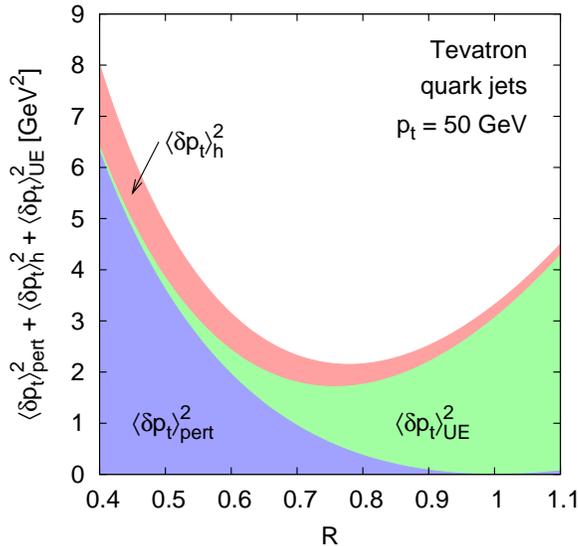}
  \caption{Contributions to the average $\delta p_t^2$ from
    perturbative radiation, hadronization and underlying event, for quark jets 
    at the Tevatron, as a function of $R$. For the perturbative contribution
    we have used \eq{eq:deltapt_pert_smallR_res}; for the hadronization
    contribution we have taken just the $1/R$ term; for UE we have used 
    the full $R$ dependence and set $\Lambda_\UE = 4 \GeV$.}
  \label{fig:sum-sq-v-r}
\end{figure}

When considering how well one can reconstruct kinematic structures
such as mass peaks, one needs to know the dispersion due to both
perturbative and non-perturbative effects, as well as any non-trivial
correlations between them. Although this goes beyond the scope of what has 
been calculated in this paper, some basic quantitative information can 
nonetheless be obtained, by arguing that the dispersion on a jet $p_t$ 
(and therefore on any kinematic structure) can be approximated
by the uncorrelated sum
\beq
  \label{eq:disp-jet-pt}
  \langle \delta p_t^2 \rangle \sim
  \langle \delta p_t \rangle^2_{\rm pert} + 
  \langle \delta p_t \rangle^2_\h + 
  \langle \delta p_t \rangle^2_\UE \, .
\eeq
This size of the summed squared contributions is shown as a function
of $R$ in Fig.~\ref{fig:sum-sq-v-r}, using our analytical results for
the perturbative and for the two non-perturbative contributions, for $50\GeV$
quark jets, with the normalisation for the underlying event
contribution set to $\Lambda_\UE = 4 \GeV$, corresponding roughly to
{\tt Pythia}'s estimate for the Tevatron. One observes that while
perturbation theory prefers\footnote{The dispersion of the
perturbative contribution will not of course precisely vanish for $R = 1$ 
--- rather, Fig.~\ref{fig:sum-sq-v-r} should be thought of as the extra
dispersion over and above whatever minimal perturbative dispersion
would be present for $R=1$.} $R \simeq 1$, the significant underlying 
event contribution leads one to favour somewhat smaller values, $R 
\simeq 0.7 - 0.8$. For this $p_t$ value, on the other hand, hadronization 
has  a relatively limited effect.

\begin{figure}[ht]
  \centering
  \includegraphics[width=0.55\textwidth]{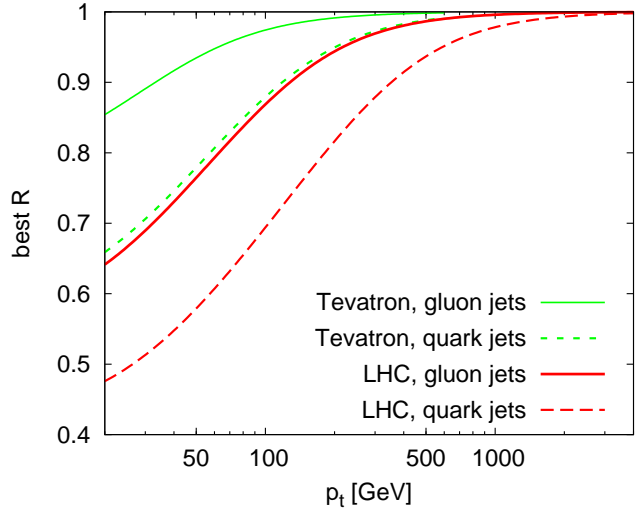}
  \caption{The $R$ value that minimises the sum of squared average
    perturbative, hadronization and UE contributions, as a function of
    $p_t$. The approximations are the same as those in
    Fig.~\ref{fig:sum-sq-v-r}, except that for LHC we have used
    $\Lambda_{UE} = 10\GeV$.}
  \label{fig:best-R-v-pt}
\end{figure}

One can make plots similar to Fig.~\ref{fig:sum-sq-v-r} for a range of
jet transverse momenta, for quark jets and gluons jets, and at
Tevatron or at LHC. In each case one can determine an optimal $R$,
minimising the combination of perturbative and non-perturbative
contributions, \eq{eq:disp-jet-pt}. The results 
are shown, as a function of $p_t$, in Fig.~\ref{fig:best-R-v-pt}.

One sees that in the high-$p_t$ limit, where perturbative radiation
dominates, $R$ should be chosen $\simeq 1$, as expected, since this 
minimises $|\ln R|$.  At lower transverse momentum, $p_t \sim 100\GeV$, 
the contribution from the UE, with its large intrinsic scale $\sim 4-10\GeV$
becomes of similar relative importance as the perturbative $\as p_t$ term, so
that it becomes advantageous to decrease $R$, in order to take advantage of
the $R^2$ reduction of the UE contribution. As a consequence, even though 
on average jets are slightly narrower at high $p_t$ than at low $p_t$ 
(essentially because of the reduction of $\alpha_s$), it is actually 
advantageous to use \emph{larger} $R$ values at high $p_t$, because in 
absolute terms perturbative radiation is roughly proportional to $p_t$.

In contrast to this slightly counterintuitive result, when going from quark 
to gluon jets one's intuition about the correlation between jet width and
choice of $R$ is reliable: for gluon jets, perturbative radiation and hadronization
are both larger, whereas the underlying event is unchanged; the optimal $R$,
therefore, is larger. One may finally note that at LHC one needs a smaller 
$R$ than at Tevatron, because of the noisier underlying event.

The situation is different if one considers observables where perturbative 
effects are accounted for by higher-order calculations. One should then 
minimise just the sum of the squares of the hadronization and UE components.
Ignoring all but the $R^2$ term of the UE contribution, this implies that the 
optimal $R$ is given by
\beq
  \label{eq:optimal-R-hadr-UE}
  R = \sqrt{2} \left( \frac{C_i {\cal A}(\mu_I)}{\Lambda_\UE} 
  \right)^{1/3}\,,
\eeq
where, as usual, $C_i$ is the colour factor for the hard parton 
responsible for the jet. Using the same values for $\Lambda_\UE$ 
as in Fig.~\ref{fig:best-R-v-pt}, this gives the results in 
Table~\ref{tab:best-R-non-pert}, which are independent of $p_t$ to a 
first approximation, and actually correspond to the zero-$p_t$ limit of 
Fig.~\ref{fig:best-R-v-pt}.
\begin{table}
  \centering
  \begin{tabular}{l|cc}
             & quark jets & gluon jets \\ \hline
    Tevatron &  0.63 & 0.83 \\
    LHC      &  0.46 & 0.61
  \end{tabular}
  \caption{$R$ values that minimise the two non-perturbative
    contributions in various circumstances.}
  \label{tab:best-R-non-pert}
\end{table}

It should be kept in mind that these results involve many approximations: 
for example, we have ignored differences between jet algorithms, we have 
used just the small $R$ limit for some terms, we have assumed that
average shifts are indicative of dispersions, and we have neglected the
issue of actually being able to resolve separate jets for complex
massive particle decays. We believe, nevertheless, that the basic conclusions 
are valid all the same: in reconstructing kinematic structures one
should prefer larger $R$ at high $p_t$ and for gluon jets, and smaller
$R$ at the LHC than at the Tevatron; when perturbative effects are
calculated separately, on the other hand, the minimisation of the
non-perturbative contributions alone leads one to favour somewhat lower
values of $R$.

\subsection[Measurement of ${\langle \delta p_t \rangle}$]{Measurement
  of $\boldsymbol{\langle \delta p_t \rangle}$} 
\label{sec:meas-delt-pt}

Throughout this article we have discussed the change in a jet transverse
momentum due to perturbative and non-perturbative effects. It should be 
emphasised, however, that this $\langle \delta p_t \rangle$ is not an 
operationally well-defined quantity away from the threshold limit.  The 
reason is that we have imagined using a parton to provide a reference 
transverse momentum, with respect to which we then discuss the shift
$\langle \delta p_t \rangle$. Beyond leading order, collinear divergences 
and quantum mechanical interference between emissions make the choice 
of a reference parton $p_t$ inherently ambiguous. Were we discussing 
$\ee$ collisions, this would not be a serious issue, since the $\ee$ centre 
of mass energy itself would provide a meaningful reference scale. At hadron colliders, however, and notably in dijet production, such an unambiguous 
reference scale seldom exists.

\begin{figure}
  \centering
  \includegraphics[width=0.6\textwidth]{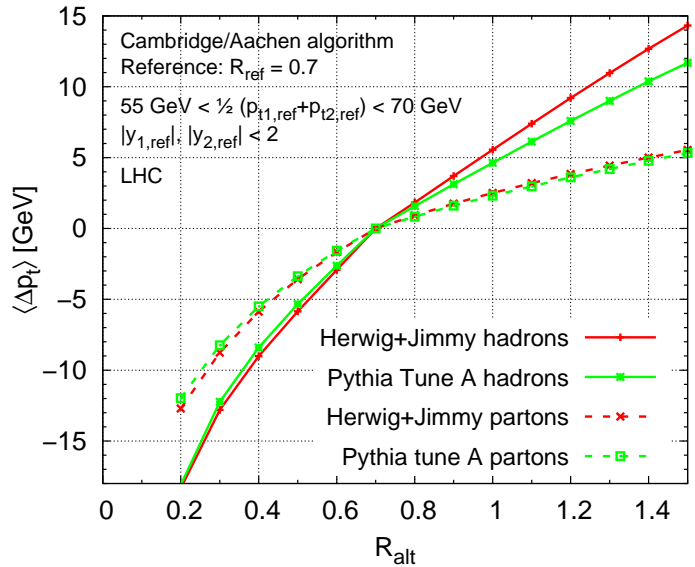}
  \caption{The average shift, $\langle \delta p_t \rangle$, as defined
    in \eq{eq:delta-pt-meas}), between an alternate (Cambridge/Aachen
    with variable $R_\alt$) and a reference jet definition (the same
    algorithm, with fixed $R_\rf = 0.7$), determined from Monte Carlo
    simulation.  Hadron level results include the UE contribution,
    whereas at parton level it has been switched off. See text for
    further details.}
  \label{fig:dpt}
\end{figure}

To study directly the shift $\langle \delta p_t \rangle$, either at NLO or
experimentally, one must therefore find a physical way of identifying a 
reference scale for the jet. One possible approach, which we propose to 
follow here, is to introduce a reference jet definition, $D_\rf$, and then 
measure how a jet transverse momentum changes if one uses an alternative
definition $D_\alt$. One complication is that different jet definitions lead, in
principle, to different sets of jets, so that one must be able to identify which 
jet, with definition $D_\alt$, corresponds to a given jet with definition
$D_\rf$. An algorithm for doing so could be outlined as follows.
\begin{enumerate}
\item Identify the set of reference jets that are of interest,
  $j_\rf^{(1)}, \ldots, j_\rf^{(n)}$, for example the two hardest
  jets in the event, using the definition $D_\rf$.
\item Associate with each jet in the reference set, $j_\rf^{(i)}$, the
  alternate jet $j_\alt^{(k)}$ with which it shares the most $p_t$.
  In other words, find the $k$ that maximises
  \beq
    \label{eq:shared-pt}
    p_t^{(k i)} = \sum_{\ell \, \in \, 
      j_\alt^{(k)} \, \cap \, j_\rf^{(i)}} \!\!\!\!  p_{t,\ell} \, \, ,
  \eeq
  where the sum runs over particles $\ell$ that are contained in both jets.
\item Reject events in which the same alternate jet is associated with 
  multiple reference jets\footnote{In the studies described below, we find 
  that this occurs only very rarely: in a fraction of a percent of events when
  using the Cambridge/Aachen algorithm both as reference and alternate, 
  and in a couple of percent of events if the alternate algorithm
  is SISCone.}, as well as events in which a reference jet $j_\rf^{(i)}$
  shares no $p_t$ with any of the $j_\rf^{(k)}$, $p_t^{(k i)} = 0 \, , \,
  \forall k$. 
\item Discard alternate jets that have not been associated with any 
  reference jets. 
\end{enumerate}
Being interested, say, in the $R$ dependence of $\langle \delta p_t 
\rangle$, one might then perform the following analysis. Begin by selecting, 
with the reference algorithm, events in which the two hardest jets are both
central, so that they are well measured (set, for example, $| y_\rf^{(i)} | 
< 2 $); require the sum of the jet $p_t$'s to be in a some limited
range, for example $55 \GeV < \frac12 (p_{t,\rf}^{(1)} + p_{t,\rf}^{(2)}) 
< 70 \GeV$; repeat the jet finding with the same jet algorithm but with an
alternate radius $R_{\alt}$; find the jets that match the two reference jets, 
and determine the difference in $p_t$ between the alternate and reference 
jet definitions, using
\beq
  \label{eq:delta-pt-meas}
  \delta p_t = \frac12 \left( p_{t,\alt}^{(1)} + p_{t,\alt}^{(2)}
    - p_{t,\rf}^{(1)} - p_{t,\rf}^{(2)}
  \right)~.
\eeq
For illustrative purposes, this procedure has been carried out on LHC
events simulated with {\tt Pythia} and with {\tt Herwig}, and the resulting 
$\langle \delta p_t \rangle$ is shown Fig.~\ref{fig:dpt}, both at parton level 
and at hadron level (including hadronization and UE). We used the
Cambridge/Aachen algorithm with reference jet radius $R_\rf =
0.7$. One notes that the moderate differences between the {\tt Pythia}
tune~A and {\tt Jimmy} underlying events (cf.\ Fig.~\ref{fig:lhc-gg}) are
clearly visible here. One also observes that perturbative and 
non-perturbative effects are of comparable sizes.

Given a measurement of $\langle \delta p_t \rangle$, an interesting
study would be to compute the average $p_t$ change at NLO, $\langle
\delta p_t \rangle_\NLO$, for example with 
{\tt NLOjet++}~\cite{Nagy:2003tz}. Such a calculation would include
configurations with up to three partons in a jet (a level of detailed jet 
structure that is reached in the inclusive jet spectrum only at NNLO). 
Given this accurate perturbative knowledge of $\delta p_t$, one could then
fit the two non-perturbative components, in order to get data-based
constraints on their magnitude independent of parton shower
Monte Carlos. Specifically, one would write\footnote{The hadronization
  terms of higher order in $R$ have been neglected here, since in
  the region of $R$ where they contribute significantly they are dominated   
  by the UE contribution.}
\beqa
  \label{eq:delta-pt-fit}
  \langle \delta p_t \rangle (R_\alt) & = & 
  \langle \delta p_t \rangle_\NLO (R_\alt)
  - 2 \, \langle C_i \rangle \left(\frac{1}{R_\alt} -
    \frac{1}{R_\rf} \right) {\cal A} (\mu_I)  \nonumber \\
  && \qquad \qquad + \, \, \left(R_\alt \, J_1(R_\alt) - R_\rf \, J_1(R_\rf) \right) 
  \Lambda_\UE \, ,
\eeqa
where $\langle C_i \rangle$ is the average colour factor that takes
into account the fraction of quark and gluon jets.
Using \eq{eq:delta-pt-fit}, one would be able to fit the two non-perturbative
parameters ${\cal A} (\mu_I)$ and $\Lambda_\UE$. One would expect 
${\cal A} (\mu_I)$ to agree qualitatively with value extracted from event 
shape studies in $e^+ e^-$ annihilation and DIS~\cite{Dasgupta:2003iq}, 
and one would obtain a first direct estimate of $\Lambda_\UE$.

A similar analysis could also be carried out in other contexts, for example
$W+$jet events. In that case one would consider only a single reference jet, 
and one could use {\tt MCFM}~\cite{Campbell:2002tg,Campbell:2003hd} for 
the perturbative calculation. Since the fraction of quark and gluon jets would 
be rather different in this case, compared to inclusive jets (especially at LHC),
this would provide a powerful check of the dependence of hadronization on 
partonic colour factors,  and it would also test whether the UE is independent 
of the underlying hard partonic reaction, as expected in many simple models.

\subsection{Inclusive jet spectrum}
\label{sec:incl-jet-spectr}

A context where non-perturbative corrections to jets have often been
discussed~\cite{Campbell:2006wx,Mangano:1999sz} is the
inclusive jet $p_t$ spectrum. In Ref.~\cite{Mangano:1999sz}, for example,
one sees that introducing a non-perturbative shift in $p_t$ is necessary in 
order to fit data for the ratio of inclusive distributions measured at UA1 and 
at Tevatron. In order to estimate the effects of the nonperturbative shift we 
have computed here, let us begin by assuming that we have at our disposal
a jet definition such that quark and gluon jets can be distinguished at parton 
level (an example of such a definition is given in~\cite{Banfi:2006hf}), and such
that it can be consistently employed at hadron level as well (which is not the 
case for the definition in~\cite{Banfi:2006hf}). Then we would be able to write
\beq
  \label{eq:dsig-dpt-flav-sep}
  \frac{d \sigma}{d p_t} \left( p_t \right) = \frac{d \sigma^{q, \PT}}{d p_t}
  \Big(p_t - \langle \delta p_t^q \rangle_{\NP} \Big)  + 
  \frac{d \sigma^{g, \PT}}{d p_t}\Big( p_t -
    \langle  \delta p_t^g \rangle_{\NP} \Big) \, ,
\eeq
where $d \sigma^{i, \PT}/d p_t$ is the perturbative distribution for jets of 
flavor $i$, which is then evaluated with an $i$-dependent shift
\beq
  \label{eq:dpt-np-incl}
  \langle  \delta p_t^i \rangle_{\NP} = 
  - 2 \, \frac{C_i}{R} {\cal A} (\mu_I) \, + \, R \, J_1(R) \, \Lambda_{UE} \, .
\eeq
It should be emphasised that, even at this stage, \eq{eq:dsig-dpt-flav-sep} 
relies on several approximations: it assumes, for example, that one can neglect 
the effect of event-by-event fluctuations in the hadronization and UE
contributions, and it assumes that non-perturbative effects are dominated
by an overall shift in the distribution. This is essentially the same approximation 
that led to the broadly successful use of a constant shift in place of a shape
function when studying non-perturbative effects in event-shape
distributions (as in~\cite{Korchemsky:1994is,Dokshitzer:1997ew,Gardi:2001ny}, see however also~\cite{Gardi:2002bg,Gardi:2003iv,Berger:2003pk,Berger:2004xf}). One should note that here, even at the level of single gluon exchange,
soft gluon resummation near threshold suggests that the distribution 
will be distorted, and not simply shifted, due to colour mixing 
effects~\cite{Dasgupta:2007hr}. It can be shown, however, that the leading
$1/R$ term is not affected by colour mixing in the context of NLL resummation.
We can then safely assume that the shift in \eq{eq:dpt-np-incl} is the 
dominant effect: the first term is large for $R < 1$, while the second term is
large because of the large value of $\Lambda_\UE$.

In practice, \eq{eq:dsig-dpt-flav-sep} is inconvenient because of our inability 
to separately define $\sigma^{q,\PT}$ and $\sigma^{g,\PT}$ beyond leading
order (LO), for jet algorithms that are valid also at hadron level. We may
however further simplify \eq{eq:dsig-dpt-flav-sep} by expanding in the small 
relative shifts $\langle \delta p_t^i \rangle_{\NP}/p_t$. To this end, let us 
define
\beq
  \label{eq:nq-ng}
  n_i \equiv - \frac{d \ln \left(d \sigma^{i, \PT}/d p_t \right)}{d \ln p_t} \, , \quad \qquad
  f_i \equiv \frac{d \sigma^{i, \PT}}{dp_t}  \left/ \, \frac{d \sigma^{\PT}}{d 
    p_t} \right. \, ,
\eeq
which represent respectively the power-law fall-off of the distribution, which 
behaves as $p_t^{-n_i}$, and the fraction of jets that are of species $i$.
We can the write
\beq
  \label{eq:dsig-dpt-flav-sep-exp}
  \frac{d \sigma}{d p_t} \left( p_t \right) = \frac{d \sigma^{q, \PT}}{d p_t}
  \left( p_t \right) \left(1 + n_q \frac{\langle \delta p_t^q 
  \rangle_{\NP}}{p_t}  + \order{ n_q^2 \frac{\langle \delta p_t^q
  \rangle_{\NP}^2}{p_t^2} }  \right)  + 
  \, \Big( q \leftrightarrow g \Big) \, ,
\eeq
We note that the neglected term in \eq{eq:dsig-dpt-flav-sep-exp} is expected 
to be of the same magnitude as terms that would arise from the replacement of
the shift with a shape function. Finally we rewrite \eq{eq:dsig-dpt-flav-sep-exp}
as
\beq
  \label{eq:dsig-dpt-flav-recomb}
  \frac{d \sigma}{d p_t} \left( p_t \right) = \frac{d \sigma^{\PT}}{d 
  p_t} \left( p_t - \frac{f_q \, n_q \, \langle \delta p_t^q \rangle_{\NP} + 
  f_g \, n_g \, \langle \delta p_t^g \rangle_{\NP}}{f_q n_q + f_g n_g} + 
  \order{n_i^2 \, \frac{\langle \delta p_t^i \rangle_{\NP}^2}{p_t}}
  \right) \, .
\eeq
In \eq{eq:dsig-dpt-flav-recomb}, one can evaluate $d \sigma^{\PT}/d p_t$ 
up to NLO (or beyond) since it is summed over jet flavors, while for the 
purpose of calculating the (small) non-perturbative shift, $f_i$ and $n_i$ 
can be evaluated at LO, where the jet flavour is unambiguous.  
Generalising \eq{eq:dsig-dpt-flav-recomb} to include not only $1/R$
hadronization terms, but also subleading powers of $R$, is a nontrivial 
task: one would have to sum not over $q$ and $g$ jets, but rather over all
different scattering channels ($qq\to qq$, $qg\to qg$, $gg \to gg$,
etc.), and furthermore the threshold approximation suggests that colour 
correlations would make it necessary to go beyond the shift approximation.
These subleading terms would therefore be quite interesting from a theoretical viewpoint, while they are unlikely to have a significant phenomenological 
impact, given the likely experimental constraints in the difficult environment
of hadron-hadron collisions. For phenomenological applications, on the other 
hand, a further and final approximation can be made on 
\eq{eq:dsig-dpt-flav-recomb}, if one assumes that quark and gluon
species have a similar power-law fall-off, $n_q \simeq n_g$. One can 
then simply write
\beq
  \label{eq:dsig-dpt-flav-recomb2}
  \frac{d \sigma}{d p_t} \left( p_t \right) \simeq \frac{d \sigma^{\PT}}{d p_t}
  \Big(p_t - \langle \delta p_t \rangle_{\NP} \Big) \, , \quad \qquad
  \langle \delta p_t^{\NP} \rangle = 
  f_q \, \langle \delta p_t^q \rangle_{\NP} + f_g \, \langle \delta
  p_t^g \rangle_{\NP} \, ,
\eeq
Again, this is can be extended, if one wishes, to separately consider all $2 
\to 2$ scattering channels, however a full treatment of the $\order{R}$
terms in the hadronization correction will probably require that the prediction 
be rephrased in the language of shape functions.

To conclude, we observe that in addition to the inclusive jet spectrum directly, 
one may also examine the relative difference between the distributions that 
emerge when using two different jet definitions, in analogy to what was done in
\secn{sec:meas-delt-pt}. One may define
\beq
  \label{eq:ref-diff-xsct}
  \rho(D_\rf, D_\alt, p_t) = \left( \frac{d \sigma^{\alt}}{d p_t} -
  \frac{d \sigma^{\rf}}{d p_t} \right) \left/ \left( \frac{d \sigma^{\rf}}{d p_t} 
  \right) \right. \, .
\eeq
Perturbatively, with current tools \cite{Nagy:2003tz}, the numerator can be
studied up to $\as^4$, because the $\as^2$ term vanishes; in other words, 
this is effectively a three-jet observable, which can be calculated to NLO, and
thus this is a quantity that can be studied to higher perturbative order than 
the inclusive jet spectrum itself (for which $\as^4$ would correspond to 
NNLO). \eq{eq:ref-diff-xsct} is similar in this respect to the difference in 
$p_t$ between alternate and reference definitions discussed in 
\secn{sec:meas-delt-pt}. The separation of $\rho$ into perturbative and 
non-perturbative parts reads 
\beqa
  \label{eq:rho-sep}
  &&  \rho(D_\rf, D_\alt, p_t) = \rho^{\PT} (D_\rf, D_\alt, p_t) \\ 
  && \quad + \, \frac{1}{p_t} \Big[ f_q \, n_q \, \big( \langle \delta 
  p_t^q \rangle_{\NP, \alt} - \langle \delta p_t^q \rangle_{\NP, \rf} \big) + 
  f_g \, n_g  \, \big( \langle \delta p_t^g \rangle_{\NP, \alt} - \langle \delta
  p_t^g \rangle_{\NP, \rf} \big) \Big] \, . \nonumber
\eeqa
Clearly, these techniques can also be applied to ratios of distributions measured
with different centre-of-mass energies (as in~\cite{Mangano:1999sz}), or 
differing with respect to other parameters. In this way one may attempt to
minimise or isolate the perturbative contribution, and get a better handle on 
our non-perturbative inputs. When these are established, one may apply the
resulting shift to the distribution itself, in order to optimise the analysis in view
of other QCD or new physics studies.

\section{Conclusions}
\label{sec:conclusions}

We have performed a first broad analysis of non-perturbative QCD effects
on jet observables at hadron colliders such as Tevatron or the LHC, examining
hadronization effects and underlying event corrections, and studying
how they compare to perturbative contributions.
We have used both analytic models, within the framework of the dispersive
approach, and different Monte Carlo tools, implementing different IRC-safe
jet algorithms, belonging to both the `cone' and `sequential recombination' 
families of jet finders. Finally, we have considered different observables, focusing mostly on the shift in the transverse momentum of a jet due to QCD radiation,
but giving results also for the jet mass and for the jet $p_t$ distribution.

Our most significant results are displayed in
\eq{eq:summary_hadronization} and \eq{eq:summary_pp}, and collected in
Table~\ref{tab:summary-of-effects}. They show that the transverse
momentum of a jet of radius $R$ is modified by hadronization effects
in a manner proportional to $1/R$, while underlying event corrections
grow like $R^2$, increasing with the jet area. A similar pattern 
holds for our analysis of the jet mass, where in general the
correction is suppressed by two powers of $R$ with respect to the
$p_t$ analysis, but with the same hierarchy between hadronization and
underlying event. As we briefly reviewed, the perturbative
$R$-dependence for the jet $p_t$ is logarithmic at small $R$. We thus
have three different functional forms for the $R$-dependence of
the three main contributions to the jet momentum. This leads to the
possibility of selectively minimising chosen combinations of these
three effects, by a suitable choice of $R$, in order to optimise a
given physics analysis.

Our results are formulated in terms of two different non-perturbative
parameters: the first, ${\cal A} (\mu_I)$, describes hadronization,
and we expect it to be closely related to similar parameters measured
in event-shape studies in $e^+ e^-$ annihilation, a conjecture which
is consistent with the Monte Carlo results; the second parameter,
$\Lambda_\UE$, describes the flow of transverse momentum per unit
rapidity due to underlying hadronic activity in a given collider
environment: its value must be extracted from data and is expected to
be universal across different observables in a given collider, while
it should scale with a power of the centre-of-mass energy. 
A further characterisation is provided by the dependence of transverse
momentum on the colour charge of the parton originating the jet: in our
approximations, both perturbative and hadronization contributions are
proportional to the colour charge, while the underlying event is
independent of it.\footnote{Work is already in progress~\cite{CCSArea}
  to understand how perturbative and UE effects interact, so as to
  introduce a subleading dependence on the jet colour factor in the
  UE contribution.}.

These result are rather strikingly confirmed by our studies with Monte Carlo simulations. Hadronization corrections, as given by non-perturbative models
embedded in {\tt Herwig} and {\tt Pythia}, indeed grow at small $R$, broadly
agreeing with the analytic estimate both in shape and normalisation, while the
corresponding underlying event corrections (computed with {\tt Jimmy} for the
{\tt Herwig} simulation) grow with $R$ as expected. 

Finally, we have given three examples of experimental studies where our 
results could be either tested or used in order to perform physics analyses.
Specifically, we have outlined how our results are likely to impact the 
single-inclusive jet $p_t$ distribution, a signature observable at hadron 
colliders in general and at LHC in particular; we have indicated how the 
$p_t$ shift from parton to hadron level could be experimentally accessed 
in order to test our calculation; and we have suggested methods and 
choices in order to optimise the value of the jet radius in view of different
experimental situations.

We regard these suggestions as just basic examples of the studies that
could be performed. More generally, we believe that our study
emphasises the necessity for variety and flexibility in the
experimental choices of jet definitions and parameters. In the early
days of QCD, when theoretical predictions were mostly at leading order
in perturbation theory, it was reasonable to adopt a rough definition
of a jet, and several such definitions proved sufficient to gather
striking evidence for many basic features of QCD and of the standard
model. Those days are past: QCD is now precision physics, with NLO
calculations forming the standard benchmark for predictions, selected
observables having been computed at higher orders, and all-order
resummations and power correction studies also available. From an
experimental point of view, the challenges of the high-luminosity,
high-energy environment of the LHC will require that all the
experience acquired with previous accelerators, and all our
theoretical knowledge, be put to good use.

We believe that in order to do that it will be necessary to take
advantage of the full flexibility offered by modern jet tools.
There are now several well-tested IRC-safe
jet algorithms, both of cone type and based on sequential
recombination. They depend on parameters, notably the radius $R$, which
provide the experimenter with handles to control the flow and impact
of perturbative and non-perturbative QCD radiation.
Exploiting the adaptability of these jet tools will be crucial
for model testing, for the validation of experimental procedures, and,
last but not least, for studies that will further our understanding of the
strong interactions.

\section*{Acknowledgements}
\label{sec:acknolwedgments}

We thank Matteo Cacciari for collaboration in the initial stages of
this project and Jon Butterworth, Arthur Moraes, Torbjorn
Sj\"ostrand and Peter Skands for discussions.
We are grateful to each other's home institution, as well as to the
Galileo Galilei Institute (Florence, Italy), and to the CERN Theory
Division for hospitality during the completion of this work. MD thanks
the CNRS for a visiting fellowship during the period in which this
collaboration was initiated.
Work supported by MIUR under contract 2006020509$\_$004, by the French
ANR under contract ANR-05-JCJC-0046-01 and by the European Community's
Marie-Curie Research Training Network `Tools and Precision
Calculations for Physics Discoveries at Colliders' (`HEPTOOLS'), under
contract MRTN-CT-2006-035505.

\appendix

\section*{Appendix}

Here we compute the global term involving the integral of the unrecombined 
gluon contribution $\delta p_t^-$, defined in \eq{delpt}, over all of phase space, 
to show that after accounting for the evolution of parton distributions it
does not contribute any linear power correction. To illustrate the point, we 
consider the dipole formed by the two incoming legs, discussed in \secn{ssec:12}. To restrict our attention to soft emission alone we impose a cutoff, requiring 
that the longitudinal fraction (with respect to the beam) of the emitted gluon 
momentum $k$ never exceed $\epsilon_c$, implying $|\eta| < \eta_{\max} 
(k_t) = \ln \epsilon_c \sqrt{s}/k_t$. The relevant integral is then 
\beq
  \label{eq:dpt-glob}
  \langle \delta p_t \rangle_{\rm gl}^{(12)} = - \, \frac{\CF}{\pi} \int
  d k_t \, \delta \as (k_t)  \, d \eta \, \frac{d \phi}{2 \pi} \,
  \big(\cosh( \eta ) + \cos \phi \big) \, .  
\eeq
Here  we have set the colour factor of the dipole $C_{12}$ to be $2 C_F$, 
since we are interested only in the radiation collinear to the incoming legs, 
which we take to be quarks. In actual fact the colour charge $C_F$ for each 
leg will be built up from a sum of dipole contributions, all of which involve the 
incoming leg in question and give the same result in the collinear limit: the replacement of the dipole colour factor by $2 C_F$ anticipates this fact.
To illustrate our point, but without loss of generality, we use a toy model for 
the non-perturbative coupling, setting $\delta \alpha_s (k_t) = \Lambda 
\delta \left(k_t - \Lambda \right)$.  In this model the quantity $\mathcal{A}$,
which governs our leading power corrections computed in \secn{sec:analytical},
is just proportional to $\Lambda$. After performing the integral over the full 
range of azimuth $\phi$, the global term gives
\beqa
\label{eq:dpt-mean}
  \langle \delta p_t \rangle_{\rm gl}^{(12)} & = & - \, \frac{\CF}{\pi} 
  \int \delta \alpha_s (k_t) \, d k_t \int_{- \eta_{\max}}^{\eta_{\max}} 
  d\eta \, \cosh \eta \nonumber \\
  & = &  - \, \frac{2 \CF}{\pi} \int \delta \alpha_s (k_t) \, d k_t 
  \sinh \eta_{\max} \\
  & = & - \, \frac{\CF}{\pi} \cdot \epsilon_c \sqrt{s} +
  \order{\frac{\Lambda^2}{\sqrt{s}}} \, . \nonumber 
\eeqa
As we shall see, the dependence on the cutoff $\epsilon_c$ will be removed by a similar term 
due to the evolution of the parton distributions. Note then the absence of any 
term of order $\Lambda$, and hence the absence of a leading power correction due to this global term.

In order to verify the factorisability of the collinear divergence in 
\eq{eq:dpt-mean}, let us now turn to the contribution of parton distributions. 
In the calculations presented in section \ref{sec:analytical} 
we worked in the threshold limit, effectively approximating parton distributions 
with their bare expression at threshold, $q_0 (x) = \delta(1 - x)$.
In our non-perturbative approximation, the distribution evolved to a hard 
scale $Q$ is then given by
\beqa
  \label{eq:PDF}
  q (x, Q^2) & = & \delta(1 - x) + \frac{2 \CF}{\pi} \int_{1 - \epsilon_c}^{1}
  \frac{d z}{1 - z} \int_{Q_0}^{Q} \frac{d k_t}{k_t} \, \delta \as(k_t) \, 
  \big( \delta(x - z) - \delta(1 - x) \big) \, , \nonumber \\
  & = &  \delta(1 - x) + \frac{2 \CF}{\pi} \, \int_{1 - \epsilon_c}^{1}
  \frac{d z}{1 - z} \left( \delta(x - z) - \delta(1 - x) \right) \, ,
\eeqa
where $Q_0$ is an arbitrary reference scale,  we have placed the same 
cutoff as in \eq{eq:dpt-mean} on the maximum emitted longitudinal 
momentum, and we have used the infrared limit ($z \to 1$) of the quark 
splitting function $P_{q q} (z)$. Note that when one considers parton 
evolution one automatically includes collinear branchings which push the 
hard scattering away from threshold.

Taking into account parton evolution, we are then forced to include a further 
non-perturbative shift in the transverse momentum of the jet, given by
\beqa
  \label{eq:jet-pdf-pt}
  \langle \delta p_t \rangle_{\rm PDF} & = &
  \int d x_1 d x_2 \frac{\sqrt{x_1 x_2 s }}{2} \, \big( q(x_1, Q^2) \, 
  q(x_2, Q^2) - q_0 (x_1) \, q_0 (x_2) \big) \nonumber \\
  & = & \frac{2\CF}{\pi} \left( 
  \int_{1 - \epsilon_c}^1 \frac{d z_1}{1 - z_1}
  \frac{(\sqrt{z_1} - 1) \sqrt{s}}{2} + (1 \leftrightarrow 2)
  \right) \\
  & = & - \, \frac{\CF}{\pi} \epsilon_c \sqrt{s} + \order{\epsilon_c^2} 
  \nonumber \, .
\eeqa
One observes that the collinear divergence in \eq{eq:dpt-mean} has precisely
the form required to be absorbed in a `renormalised' parton distribution, such 
as the one computed in \eq{eq:jet-pdf-pt}. \eq{eq:dpt-mean} then assures us 
that the remaining finite terms do not contain leading power corrections.

Similar calculations can be carried out for all dipoles,  and in each case one
observes that no ${\mathcal{O}}(\Lambda)$ term arises from `global'
integrations over all phase space. For each dipole, furthermore, collinear
divergences can either be factorised into the parton distribution, as done above,
or they cancel, as in the case of outgoing legs, against an identical divergence 
in the 'in-jet' contribution, as expected from the collinear safety of our 
observables.


\begin{thebibliography}{99}

\bibitem{Kidonakis:1997gm}
  N.~Kidonakis and G.~Sterman,
  Nucl.\ Phys.\  B {\bf 505} (1997) 321,
  {\tt hep-ph/9705234}.

\bibitem{Kidonakis:1998bk}
  N.~Kidonakis, G.~Oderda and G.~Sterman,
  Nucl.\ Phys.\  B {\bf 525} (1998) 299,
  {\tt hep-ph/9801268}.

\bibitem{Kidonakis:1998nf}
  N.~Kidonakis, G.~Oderda and G.~Sterman,
  Nucl.\ Phys.\  B {\bf 531} (1998) 365,
  {\tt hep-ph/9803241}.

\bibitem{Kidonakis:2000gi}
  N.~Kidonakis and J.~F.~Owens,
  Phys.\ Rev.\  D {\bf 63} (2001) 054019,
  {\tt hep-ph/0007268}.

\bibitem{deFlorian:2007fv}
  D.~de Florian and W.~Vogelsang,
  Phys.\ Rev.\  D {\bf 76} (2007) 074031,
  {\tt arXiv:0704.1677 [hep-ph]}.

\bibitem{Seymour:1997kj}
  M.~H.~Seymour,
  Nucl.\ Phys.\  B {\bf 513} (1998) 269,
  {\tt hep-ph/9707338}.

\bibitem{Korchemsky:1994is}
  G.~P.~Korchemsky and G.~Sterman,
  Nucl.\ Phys.\  B {\bf 437} (1995) 415,
  {\tt hep-ph/9411211}.

\bibitem{Dokshitzer:1995zt}
  Y.~L.~Dokshitzer and B.~R.~Webber,
  Phys.\ Lett.\  B {\bf 352} (1995) 451,
  {\tt hep-ph/9504219}.

\bibitem{Ball:1995ni}
  P.~Ball, M.~Beneke and V.~M.~Braun,
  Nucl.\ Phys.\  B {\bf 452} (1995) 563,
  {\tt hep-ph/9502300}.

\bibitem{Dokshitzer:1995qm}
  Y.~L.~Dokshitzer, G.~Marchesini and B.~R.~Webber,
  Nucl.\ Phys.\  B {\bf 469} (1996) 93,
  {\tt hep-ph/9512336}.

\bibitem{Gardi:2001di}
  E.~Gardi,
  Nucl.\ Phys.\  B {\bf 622} (2002) 365,
  {\tt hep-ph/0108222}.

\bibitem{Beneke:1998ui}
  M.~Beneke,
  Phys.\ Rept.\  {\bf 317} (1999) 1,
  {\tt hep-ph/9807443}.

\bibitem{Dasgupta:2003iq}
  M.~Dasgupta and G.~P.~Salam,
  J.\ Phys.\ G {\bf 30} (2004) R143,
  {\tt hep-ph/0312283}.

\bibitem{Banfi:2001aq}
  A.~Banfi, G.~Marchesini, G.~Smye and G.~Zanderighi,
  JHEP {\bf 0108}, 047 (2001),
  {\tt hep-ph/0106278}.

\bibitem{Dasgupta:2007hr}
  M.~Dasgupta and Y.~Delenda,
  {\tt arXiv:0709.3309 [hep-ph]}.

\bibitem{Bhatti:2005ai}
  A.~Bhatti {\it et al.},  Nucl.  Instrum.  Meth.   {\bf A 566} 375 (2006),
  {\tt hep-ex/0510047}.

\bibitem{Mangano:1999sz}
  M.~L.~Mangano,
  {\tt hep-ph/9911256}.

\bibitem{Campbell:2006wx}
  J.~M.~Campbell, J.~W.~Huston and W.~J.~Stirling,
  Rept.\ Prog.\ Phys.\  {\bf 70} (2007) 89,
  {\tt hep-ph/0611148}.

\bibitem{Ellis:2007ib}
S.~D.~Ellis, J.~Huston, K.~Hatakeyama, P.~Loch and M.~Toennesmann,  
  {\tt arXiv:0712.2447 [hep-ph]}.  

\bibitem{Sjostrand:2006za}
  T.~Sjostrand, S.~Mrenna and P.~Skands,
  JHEP {\bf 0605} (2006) 026,
  {\tt hep-ph/0603175}.

\bibitem{Corcella:2002jc}
  G.~Corcella {\it et al.},
  {\tt hep-ph/0210213}.

\bibitem{Butterworth:1996zw}
  J.~M.~Butterworth, J.~R.~Forshaw and M.~H.~Seymour,
  Z.\ Phys.\  C {\bf 72} (1996) 637,
  {\tt hep-ph/9601371}.

\bibitem{Lee:2006nr}
  C.~Lee and G.~Sterman,
  Phys.\ Rev.\  D {\bf 75} (2007) 014022,
  {\tt hep-ph/0611061}.

\bibitem{Furman:1981kf}
  M.~Furman,
  Nucl.\ Phys.\  B {\bf 197} (1982) 413.

\bibitem{Aversa:1988vb}
  F.~Aversa, P.~Chiappetta, M.~Greco and J.~P.~Guillet,
  Nucl.\ Phys.\  B {\bf 327} (1989) 105;
  Z.\ Phys.\  C {\bf 46} (1990) 253.

\bibitem{Guillet:1990ez}
  J.~P.~Guillet,
  Z.\ Phys.\  C {\bf 51} (1991) 587.

\bibitem{Jager:2004jh}
  B.~Jager, M.~Stratmann and W.~Vogelsang,
  Phys.\ Rev.\  D {\bf 70} (2004) 034010,
  {\tt hep-ph/0404057}.

\bibitem{inclukt}
  S.~Catani, Y.~L.~Dokshitzer, M.~H.~Seymour and B.~R.~Webber,
  Nucl.\ Phys.\ B {\bf 406}  (1993)  187 and refs.\ therein.

\bibitem{ESkt}
  S.~D.~Ellis and D.~E.~Soper,
  Phys.\ Rev.\ D {\bf 48} (1993) 3160,
  {\tt hep-ph/9305266}. 

\bibitem{Caachen}
  Y.~L.~Dokshitzer, G.~D.~Leder, S.~Moretti and B.~R.~Webber,
  JHEP {\bf 9708}, 001 (1997),
  {\tt hep-ph/9707323};
  M.~Wobisch and T.~Wengler,
  {\tt hep-ph/9907280}.
  
\bibitem{Salam:2007xv}
  G.~P.~Salam and G.~Soyez,
  JHEP {\bf 0705} (2007) 086,
  {\tt arXiv:0704.0292 [hep-ph]}.

\bibitem{Blazey}
  G.~C.~Blazey {\it et al.},
  {\tt hep-ex/0005012}.

\bibitem{Catani:1990rr}
  S.~Catani, B.~R.~Webber and G.~Marchesini,
  Nucl.\ Phys.\  B {\bf 349} (1991) 635.

\bibitem{Ellis:1991qj}
  R.~K.~Ellis, W.~J.~Stirling and B.~R.~Webber,
  Camb.\ Monogr.\ Part.\ Phys.\ Nucl.\ Phys.\ Cosmol.\  {\bf 8} (1996) 1.

\bibitem{Dokshitzer:1997iz}
  Y.~L.~Dokshitzer, A.~Lucenti, G.~Marchesini and G.~P.~Salam, 
  Nucl.\ Phys.\ B {\bf 511} (1998) 396
  [Erratum-ibid.\ B {\bf 593} (2001) 729],
  {\tt hep-ph/9707532}.

\bibitem{Dokshitzer:1998pt}
  Y.~L.~Dokshitzer, A.~Lucenti, G.~Marchesini and G.~P.~Salam,
  JHEP {\bf 9805} (1998) 003,
  {\tt hep-ph/9802381}.

\bibitem{Dasgupta:1998xt}
  M.~Dasgupta and B.~R.~Webber,
  JHEP {\bf 9810} (1998) 001,
  {\tt hep-ph/9809247}.

\bibitem{Dasgupta:1999mb}
  M.~Dasgupta, L.~Magnea and G.~Smye,
  JHEP {\bf 9911} (1999) 025,
  {\tt hep-ph/9911316}.

\bibitem{Smye:2001gq}
  G.~E.~Smye,
  JHEP {\bf 0105} (2001) 005,
  {\tt hep-ph/0101323}.

\bibitem{antikt}
  M.~Cacciari, G.~P.~Salam and G.~Soyez, LPTHE-07-03, in preparation.

\bibitem{Cacciari:2007fd}
  M.~Cacciari and G.~P.~Salam,
  {\tt arXiv:0707.1378 [hep-ph]}.

\bibitem{CCSArea}
  M.~Cacciari, G.~P.~Salam and G.~Soyez, LPTHE-07-02, in preparation.

\bibitem{Albrow:2006rt}
  M.~G.~Albrow {\it et al.}  [TeV4LHC QCD Working Group],
  {\tt hep-ph/0610012}.

\bibitem{D0kt}
  V.~M.~Abazov {\it et al.}  [D0 Collaboration],
  Phys.\ Rev.\  D {\bf 65} (2002) 052008,
  {\tt hep-ex/0108054}.

\bibitem{CDFkt}
  A.~Abulencia {\it et al.}  [CDF - Run II Collaboration],
  Phys.\ Rev.\  D {\bf 75} (2007) 092006
  [Erratum-ibid.\  D {\bf 75} (2007) 119901],
  {\tt hep-ex/0701051};
 A.~Abulencia {\it et al.}  [CDF II Collaboration],
  Phys.\ Rev.\ Lett.\  {\bf 96} (2006) 122001,
  {\tt hep-ex/0512062}.

\bibitem{Nagy:2003tz}
  Z.~Nagy,
  Phys.\ Rev.\  D {\bf 68} (2003) 094002,
  {\tt hep-ph/0307268}.

\bibitem{Campbell:2002tg}
  J.~Campbell and R.~K.~Ellis,
  Phys.\ Rev.\  D {\bf 65} (2002) 113007,
  {\tt hep-ph/0202176}.

\bibitem{Campbell:2003hd}
  J.~Campbell, R.~K.~Ellis and D.~L.~Rainwater,
  Phys.\ Rev.\  D {\bf 68} (2003) 094021,
  {\tt hep-ph/0308195}.

\bibitem{Banfi:2006hf}
  A.~Banfi, G.~P.~Salam and G.~Zanderighi,
  Eur.\ Phys.\ J.\  C {\bf 47} (2006) 113,
  {\tt hep-ph/0601139}.

\bibitem{Dokshitzer:1997ew}
  Y.~L.~Dokshitzer and B.~R.~Webber,
  Phys.\ Lett.\  B {\bf 404} (1997) 321,
  {\tt hep-ph/9704298}.

\bibitem{Gardi:2001ny}
  E.~Gardi and J.~Rathsman,
  Nucl.\ Phys.\  B {\bf 609} (2001) 123,
  {\tt hep-ph/0103217}.

\bibitem{Gardi:2002bg}
  E.~Gardi and J.~Rathsman,
  Nucl.\ Phys.\  B {\bf 638} (2002) 243,
  {\tt hep-ph/0201019}.
  
\bibitem{Gardi:2003iv}
  E.~Gardi and L.~Magnea,
  JHEP {\bf 0308} (2003) 030,
  {\tt hep-ph/0306094}.

\bibitem{Berger:2003pk}
  C.~F.~Berger and G.~Sterman,
  JHEP {\bf 0309} (2003) 058,
  {\tt hep-ph/0307394}.

\bibitem{Berger:2004xf}
  C.~F.~Berger and L.~Magnea,
  Phys.\ Rev.\  D {\bf 70} (2004) 094010,
  {\tt hep-ph/0407024}.

\end{thebibliography}
\end{document}